\documentclass[%
superscriptaddress,
nofootinbib,
twocolumn,
amsmath,amssymb,
aps,
prd,
a4paper
]{revtex4-2}

\usepackage[colorlinks=true,linkcolor=blue,citecolor=blue,linktocpage=true]{hyperref}

\usepackage{newtxtext,newtxmath}
\usepackage{graphicx}
\usepackage{dcolumn}
\usepackage{bm}
\usepackage{enumitem}

\begin{document}

\preprint{}

\title{Unified Lagrangian Framework for Galaxy Clustering: Consistent Modeling of Bias, Redshift-Space Distortions, and Reconstruction}

\author{Naonori Sugiyama}
\email{nao.s.sugiyama@gmail.com}
\affiliation{Independent Researcher, Tokyo, Japan}
\affiliation{National Astronomical Observatory of Japan, Mitaka, Tokyo 181-8588, Japan}
\thanks{Special Visiting Researcher (non-salaried)}

\date{\today}

\begin{abstract}
We present \emph{Unified Lagrangian Perturbation Theory} (ULPT), a perturbative framework for consistently modeling galaxy density fluctuations across real space, redshift space, and post-reconstruction fields. Unlike existing approaches that treat these cases separately, ULPT provides a single theoretical structure that incorporates the three essential coordinate mappings: the Lagrangian-to-Eulerian transformation, the real-to-redshift mapping induced by peculiar velocities, and the remapping from pre to post reconstruction. A key feature of our formulation is the explicit decomposition of the density field into two physically distinct components: the \emph{Jacobian deviation}, which encodes intrinsic linear and nonlinear growth, and the \emph{displacement-mapping effect}, which captures large-scale convective distortions. This separation enables a fully analytic and infrared (IR)-safe resummation, ensuring exact IR cancellation, a consistent Gaussian description of baryon acoustic oscillation (BAO) damping, and the correct residual structure in cross spectra between fields with distinct IR behavior. The perturbative expansion of ULPT naturally generates Galileon-type operators, thereby providing a compact and physically motivated operator basis for nonlinear and nonlocal Lagrangian bias, and allowing for a renormalization-free treatment of biased tracers. Within this framework, we derive a unified expression for the power spectrum that applies equally to dark matter, biased tracers, redshift-space distortions, and reconstructed fields. ULPT thus offers a robust and extensible foundation for precision modeling of large-scale structure, with potential extensions to higher-order statistics, such as the bispectrum, and to other two-point observables, such as galaxy--galaxy lensing.
\end{abstract}
\maketitle

\section{Introduction}

The spatial distribution of galaxies, which traces the large-scale structure (LSS) of the Universe, is one of the most sensitive probes in modern cosmology. Next-generation wide-field spectroscopic surveys, including the Prime Focus Spectrograph (PFS)~\cite{PFSTeam:2012fqu}, Euclid~\citep{EUCLID:2011zbd}, and the Dark Energy Spectroscopic Instrument (DESI)~\cite{DESI:2016fyo}, are now under way and have already begun to deliver three-dimensional maps of the cosmic web with unprecedented precision.

Extracting the full cosmological information from these data demands an accurate theoretical prediction of galaxy clustering. This, in turn, requires a consistent treatment of three coordinate transformations that intervene between first principles and the observations:%
\begin{enumerate}
    \item Initial-to-late-time mapping from Lagrangian coordinates to present-day Eulerian positions~\cite{Zeldovich:1969sb}.
    \item Real-to-redshift space mapping, which encodes redshift-space distortions (RSD)~\cite{Kaiser:1987qv}.
    \item Pre- to post-reconstruction mapping, introduced by the density-field reconstruction routinely applied to baryon acoustic oscillation (BAO) analyses~\cite{Sunyaev:1970eu,Peebles:1970ag,Eisenstein:2006nk}.
\end{enumerate}%
All three transformations describe how the galaxy number-density field changes when galaxy positions are displaced while conserving number. 

Each transformation has been studied in detail. Lagrangian perturbation theory (LPT), exemplified by the Zel'dovich approximation (ZA), takes the displacement vector as its fundamental variable, enabling an efficient resummation of infrared (IR) nonlinear contributions and, consequently, an accurate description of the nonlinear broadening of the BAO feature~\cite{Zeldovich:1969sb,Eisenstein:2006nj,Crocce:2007dt,Matsubara:2007wj,Carlson:2012bu,Wang:2013hwa,Sugiyama:2013gza,Senatore:2014via,Baldauf:2015xfa,Blas:2016sfa,Senatore:2017pbn,Ivanov:2018gjr,Lewandowski:2018ywf,Sugiyama:2020uil}. A rich literature also addresses the modeling of RSD~\cite{Crocce:2005xy,Matsubara:2007wj,Taruya:2010mx,Baumann:2010tm,Carrasco:2012cv,Carlson:2012bu,Wang:2013hwa} beyond standard perturbation theory (SPT)~\cite{Bernardeau:2001qr}. Moreover, analyses of the post-reconstruction density field have been shown that it not only tighten constraints on BAO but also on other cosmological parameters unrelated to BAO~\cite{Hikage:2020fte,Wang:2022nlx}, and substantial progress has been made in modeling the reconstructed density field~\cite{Hikage:2017tmm,Hikage:2019ihj,Hikage:2020fte,Sugiyama:2024eye,Sugiyama:2024ggt,Chen:2024tfp,Zhao:2024xit,Zhang:2025jef}.

At first glance, it may appear that LPT is sufficient to provide a consistent treatment of all three coordinate mappings. However, in practice, the term ``LPT'' encompasses a variety of methods rather than a unique formulation, and existing LPT-based approaches exhibit significant differences in their theoretical frameworks and approximations. Among these, the Lagrangian resummation theory (LRT)~\cite{Matsubara:2007wj}, which resums the exponential damping induced by large-scale displacements, is closely related to the Eulerian ``$\Gamma$-expansion''~\cite{Bernardeau:2008fa}. Both approaches, however, violate the IR cancellation that is expected to hold in the IR limit, resulting in an unphysical exponential suppression of power across all scales in the power spectrum.

Other extensions based on ZA that incorporate higher-order displacement vectors have also been proposed, most notably the convolution Lagrangian perturbation theory (CLPT)~\cite{Carlson:2012bu} (see also~\cite{Sugiyama:2013mpa}). Nevertheless, when applied to the reconstructed density field, these methods encounter nontrivial issues. For instance, a naive application of ZA to the reconstructed field predicts two distinct Gaussian-form exponential damping factors for modeling the nonlinear smearing of the BAO feature in the power spectrum~\cite{Padmanabhan:2008dd,Seo:2015eyw,White:2015eaa,Chen:2019lpf}. However, when IR effects are treated consistently, it has been shown mathematically that a single Gaussian-form exponential damping factor suffices, even in the post-reconstruction case, as in the pre-reconstruction scenario~\cite{Sugiyama:2024eye,Chen:2024tfp}.

These considerations highlight a fundamental limitation in current theoretical modeling: a fully consistent Lagrangian framework that remains applicable both before and after reconstruction and satisfies all theoretical requirements expected of IR effects---that is, an IR-safe formulation---has yet to be established. To fill this gap, we develop the \emph{Unified Lagrangian Perturbation Theory} (ULPT): a unified perturbative framework that consistently treats all three mappings in a single theoretical structure.

In ULPT, we reformulate LPT such that the Jacobian structure is consistently preserved even after defining density fluctuations with respect to a random reference field. The resulting framework enables a non-perturbative treatment of large-scale displacements, thereby providing three key theoretical benefits related to IR effects:%
\begin{enumerate}%

    \item[1.] Exact nonperturbative implementation of \textit{IR cancellation}, in which the contribution from large-scale displacements vanishes in the IR limit~\cite{Jain:1995kx,Scoccimarro:1995if,Kehagias:2013yd,Peloso:2013zw,Sugiyama:2013pwa,Sugiyama:2013gza,Blas:2013bpa,Blas:2015qsi,Lewandowski:2017kes}.

    \item[2.] Natural inclusion of IR-resummed models based on the wiggle-no-wiggle decomposition, which accurately describe the BAO damping feature and remain valid both before and after reconstruction~\cite{Sugiyama:2013gza,Senatore:2014via,Baldauf:2015xfa,Blas:2016sfa,Senatore:2017pbn,Ivanov:2018gjr,Lewandowski:2018ywf,Sugiyama:2020uil,Sugiyama:2024eye}.
   
    \item[3.] Correct prediction of the exponential damping that appears in the cross-power spectrum between pre- and post-reconstruction fields~\cite{Wang:2022nlx,Sugiyama:2024eye}.

\end{enumerate}%
Furthermore, ULPT is expected to offer additional advantages when comparing with observational data, particularly in capturing the shape of the power spectrum, which is an essential aspect of data analysis. While the above three benefits are theoretically well-established within our formulation, we also note three additional potential advantages that warrant further investigation: 
\begin{enumerate}

    \item[4.] The natural emergence of Galileon-type operators~\cite{Chan:2012jj,Assassi:2014fva} in our formulation may open the way for an alternative modeling of Lagrangian galaxy bias, in which bias contributions are systematically constructed from a Galileon-based operator basis.

    \item[5.] It may be capable of describing shape modifications induced by RSD, given its close connection to existing RSD models.

    \item[6.] It also appears to be a promising candidate for modeling reconstruction-induced shape changes beyond the scope of SPT, similarly to how RSD effects are treated.

\end{enumerate}

This paper is organized as follows. In Sec.~\ref{sec:LEM}, we develop the core structure of ULPT, introducing the decomposition of the density field into the Jacobian deviation and the displacement-mapping effect. In Sec.~\ref{sec:GB}, we extend this framework to model galaxy bias in a consistent manner. Sec.~\ref{sec:RRM} presents the ULPT-based formulation of RSD, while Sec.~\ref{sec:PRM} describes how reconstruction is incorporated into the same framework. In Sec.~\ref{sec:PTF}, we perform a perturbative expansion of the ULPT expressions, focusing on dark matter in real space, and identify the Galileon basis underlying the intrinsic nonlinear structure of the density field. Sec.~\ref{sec:power-spectrum} provides a unified derivation of the power spectrum in real, redshift, and reconstructed space, demonstrating the exact realization of IR cancellation and the emergence of IR-resummed models. In Sec.~\ref{sec:FP}, we discuss the implications of our formulation for the shape of the power spectrum and outline possible extensions to higher-order statistics and other observables. We conclude in Sec.~\ref{sec:Conclusion} with a summary and prospects for future work.

\section{Lagrangian-to-Eulerian Mapping}
\label{sec:LEM}

In this section, we begin the formal development of the Unified Lagrangian Perturbation Theory (ULPT). To illustrate the core structure of ULPT in its simplest setting, we consider the case of the dark matter density field in real space. 

Throughout this paper, we suppress the explicit time (or redshift) dependence of physical quantities for notational clarity, unless stated otherwise. Let $\rho(\boldsymbol{x})$ denote the dark matter density field in Eulerian coordinates, and $\bar{\rho}$ its background value. The density contrast $\delta(\boldsymbol{x})$ is then defined as
\begin{equation}
\rho(\boldsymbol{x}) = \bar{\rho} \left[1 + \delta(\boldsymbol{x}) \right].
\label{eq:delta_def}
\end{equation}

\subsection{Standard LPT Formulation}

In LPT, the mapping between the Eulerian coordinate $\boldsymbol{x}$ and the initial Lagrangian coordinate $\boldsymbol{q}$ is given by
\begin{equation}
\boldsymbol{x} = \boldsymbol{q} + \boldsymbol{\Psi}(\boldsymbol{q}),
\label{eq:displacement}
\end{equation}
where $\boldsymbol{\Psi}(\boldsymbol{q})$ is the displacement vector that describes the trajectory of a dark matter particle from its initial position $\boldsymbol{q}$ to its final position $\boldsymbol{x}$.

Assuming mass conservation, the total number of dark matter particles must remain constant under this transformation. This leads to the condition
\begin{equation}
    \rho(\boldsymbol{x})\, d^3 x = \bar{\rho}\, d^3q.
\label{eq:mass_conservation}
\end{equation}
The Jacobian determinant associated with the mapping is defined by
\begin{equation}
J(\boldsymbol{q}) = \det\left( \frac{\partial \boldsymbol{x}}{\partial \boldsymbol{q}} \right),
\label{eq:jacobian}
\end{equation}
which implies that the volume element transforms as
\begin{equation}
    d^3x = J(\boldsymbol{q})\, d^3q.
\label{eq:volume_element}
\end{equation}

Substituting Eq.~\eqref{eq:volume_element} into Eq.~\eqref{eq:mass_conservation}, we obtain
\begin{equation}
\rho(\boldsymbol{x}) = \frac{\bar{\rho}}{J(\boldsymbol{q})}.
\label{eq:rho}
\end{equation}
Accordingly, the density contrast is given by
\begin{equation}
1 + \delta(\boldsymbol{x}) = \frac{1}{J(\boldsymbol{q})}.
\label{eq:contrast}
\end{equation}

The Jacobian determinant \( J(\boldsymbol{q}) \) can be explicitly written in terms of the displacement vector as
\begin{align}
J(\boldsymbol{q}) =\
& 1 + \Psi_{i,i}(\boldsymbol{q}) \nonumber \\
& + \frac{1}{2} \left[
    \Psi_{i,i}(\boldsymbol{q}) \Psi_{j,j}(\boldsymbol{q})
    - \Psi_{i,j}(\boldsymbol{q}) \Psi_{j,i}(\boldsymbol{q})
  \right] \nonumber \\
& + \frac{1}{6} \epsilon_{ijk} \epsilon_{lmn}
    \Psi_{i,l}(\boldsymbol{q})
    \Psi_{j,m}(\boldsymbol{q})
    \Psi_{k,n}(\boldsymbol{q}).
    \label{eq:j_psi}
\end{align}
Here, indices \( i, j, k, l, m, n \) run over the Cartesian coordinates \( x, y, z \), and summation over repeated indices is assumed. \( \epsilon_{ijk} \) denotes the Levi-Civita symbol in three dimensions. \( \Psi_i \) denotes the \( i \)-th component of the displacement vector, and \( \Psi_{i,j} \equiv \partial \Psi_i / \partial q_j \) is its partial derivative with respect to the Lagrangian coordinate \( q_j \).

To derive an equivalent representation to Eq.~\eqref{eq:contrast}, we begin with the identity
\begin{equation}
    \rho(\boldsymbol{x}) = \int d^3x' \, \rho(\boldsymbol{x}') \, \delta_{\rm D}(\boldsymbol{x} - \boldsymbol{x}'),
\label{eq:delta_identity}
\end{equation}
where \(\delta_{\rm D}\) denotes the three-dimensional Dirac delta function.

Next, we change variables in the integrand by expressing \(\boldsymbol{x}'\) using the Lagrangian coordinate via Eq.~\eqref{eq:displacement}. Using the mass conservation relation in Eq.~\eqref{eq:mass_conservation}, we obtain
\begin{equation}
    \rho(\boldsymbol{x}) = \bar{\rho} \int d^3q \, \delta_{\rm D}(\boldsymbol{x} - \boldsymbol{q} - \boldsymbol{\Psi}(\boldsymbol{q})).
\label{eq:rho_delta_mapping}
\end{equation}
Dividing both sides by \(\bar{\rho}\), we arrive at an expression for the density contrast:
\begin{equation}
    1 + \delta(\boldsymbol{x}) = \int d^3q \, \delta_{\rm D}(\boldsymbol{x} - \boldsymbol{q} - \boldsymbol{\Psi}(\boldsymbol{q})).
\label{eq:delta_position}
\end{equation}

The unity on the left-hand side in Eq.~\eqref{eq:delta_position} corresponds to the background number density. If we ignore the displacement field by setting \(\boldsymbol{\Psi} = 0\), we recover
\begin{equation}
    1 = \int d^3q \, \delta_{\rm D}(\boldsymbol{x} - \boldsymbol{q}),
\label{eq:background_density}
\end{equation}
and thus the density contrast can be written as
\begin{equation}
    \delta(\boldsymbol{x}) = \int d^3q \left[ \delta_{\rm D}(\boldsymbol{x} - \boldsymbol{q} - \boldsymbol{\Psi}(\boldsymbol{q})) - \delta_{\rm D}(\boldsymbol{x} - \boldsymbol{q}) \right].
\label{eq:delta_difference}
\end{equation}

Taking the Fourier transform of Eq.~\eqref{eq:delta_difference}, we obtain
\begin{equation}
    \widetilde{\delta}(\boldsymbol{k}) = \int d^3q\, e^{-i \boldsymbol{k} \cdot \boldsymbol{q}} \left( e^{-i \boldsymbol{k} \cdot \boldsymbol{\Psi}(\boldsymbol{q})} - 1 \right),
\label{eq:delta_fourier}
\end{equation}
where, throughout this paper, we denote Fourier-transformed quantities with a tilde.

This representation is particularly useful because the density contrast is expressed entirely in terms of the displacement vector \(\boldsymbol{\Psi}\), which appears in the exponent. The exponential form simplifies the perturbative expansion and makes the expression mathematically tractable. For this reason, it has been widely adopted in analytical calculations based on LPT~\cite{Zeldovich:1969sb,Matsubara:2007wj,Carlson:2012bu,Wang:2013hwa,Sugiyama:2013mpa}.

\subsection{ULPT Formulation in Real Space}
\label{sec:core_lag}

We now present the central expression of ULPT, which provides an alternative formulation of the density contrast. Unlike the standard LPT expression given in Eq.~\eqref{eq:delta_difference}, which implicitly assumes a Lagrangian definition of the background, our formulation begins with a fully Eulerian treatment of the background density contribution. Specifically, we start from the identity
\begin{eqnarray}
    1 = \int d^3x'\, \delta_{\rm D}(\boldsymbol{x} - \boldsymbol{x}'),
\label{eq:unity_euler}
\end{eqnarray}
and substitute this into Eq.~\eqref{eq:delta_identity}, allowing us to begin directly from an identity for the density contrast $\delta$, rather than the density field $\rho$:
\begin{equation}
    \delta(\boldsymbol{x}) = \int d^3 x' \, \delta(\boldsymbol{x}') \, \delta_{\mathrm{D}}(\boldsymbol{x} - \boldsymbol{x}').
\end{equation}

We then express the integrand in terms of Lagrangian coordinates by applying the coordinate mapping and volume element transformation given in Eqs.~\eqref{eq:displacement} and~\eqref{eq:volume_element}. This leads to
\begin{equation}
    \delta(\boldsymbol{x}) = \int d^3 q \, J(\boldsymbol{q}) \, \delta(\boldsymbol{q} + \boldsymbol{\Psi}(\boldsymbol{q})) \, \delta_{\mathrm{D}}(\boldsymbol{x} - \boldsymbol{q} - \boldsymbol{\Psi}(\boldsymbol{q})).
    \label{eq:delta_rewritten}
\end{equation}

We define the \textit{Jacobian deviation}, denoted by \( \delta_{\rm J}(\boldsymbol{q}) \), as
\begin{align}
    \delta_{\rm J}(\boldsymbol{q}) & \equiv J(\boldsymbol{q}) \, \delta(\boldsymbol{q} + \boldsymbol{\Psi}(\boldsymbol{q})) 
    \nonumber \\
    & = J(\boldsymbol{q})\, \left( \frac{1}{J(\boldsymbol{q})}-1 \right) \nonumber \\
    & = 1 - J(\boldsymbol{q}),
    \label{eq:jacobian_deviation}
\end{align}
which captures the deviation of the volume element from its unperturbed value. Using Eq.~\eqref{eq:j_psi}, it can be expanded in terms of derivatives of the displacement field as
\begin{align}
    \delta_{\rm J}(\boldsymbol{q}) =\ &
    - \Psi_{i,i}(\boldsymbol{q}) \nonumber \\
    & - \frac{1}{2} \left[
        \Psi_{i,i}(\boldsymbol{q}) \Psi_{j,j}(\boldsymbol{q})
        - \Psi_{i,j}(\boldsymbol{q}) \Psi_{j,i}(\boldsymbol{q})
      \right] \nonumber \\
    & - \frac{1}{6} \epsilon_{ijk} \epsilon_{lmn}
        \Psi_{i,l}(\boldsymbol{q})
        \Psi_{j,m}(\boldsymbol{q})
        \Psi_{k,n}(\boldsymbol{q}).
\label{eq:dJ}
\end{align}

Substituting Eq.~(\ref{eq:jacobian_deviation}) into Eq.~(\ref{eq:delta_rewritten}), we obtain a compact expression for the density contrast:
\begin{equation}
    \delta(\boldsymbol{x}) = \int d^3 q \, \delta_{\rm J}(\boldsymbol{q}) \, \delta_{\mathrm{D}}(\boldsymbol{x} - \boldsymbol{q} - \boldsymbol{\Psi}(\boldsymbol{q})).
\label{eq:delta_from_j}
\end{equation}
Taking the Fourier transform yields
\begin{equation}
    \widetilde{\delta}(\boldsymbol{k}) = \int d^3 q \, e^{-i\boldsymbol{k}\cdot\boldsymbol{q}} \, e^{-i\boldsymbol{k}\cdot\boldsymbol{\Psi}(\boldsymbol{q})} \, \delta_{\rm J}(\boldsymbol{q}).
    \label{eq:delta_fourier_from_j}
\end{equation}
These expressions in Eqs.~\eqref{eq:delta_from_j} and~\eqref{eq:delta_fourier_from_j} form the basis of ULPT and underlie all subsequent calculations.

\medskip

The key distinction between Eq.~(\ref{eq:delta_from_j}) and the conventional LPT expression in Eq.~\eqref{eq:delta_difference} lies in the treatment of the background density. In Eq.~\eqref{eq:delta_difference}, the background is implicitly defined in Lagrangian coordinates through Eq.~\eqref{eq:background_density}, whereas Eq.~(\ref{eq:delta_from_j}) adopts an Eulerian perspective throughout.

This seemingly minor change has significant implications: even a uniform background density field, when first described in Eulerian coordinates and then mapped to Lagrangian coordinates, acquires a nontrivial perturbative structure through both the displacement field \( \boldsymbol{\Psi} \) and the Jacobian determinant \( J \). In particular, Eq.~\eqref{eq:unity_euler} becomes
\begin{equation}
    1 = \int d^3 q \, J(\boldsymbol{q}) \, \delta_{\mathrm{D}} \left( \boldsymbol{x} - \boldsymbol{q} - \boldsymbol{\Psi}(\boldsymbol{q}) \right).
    \label{eq:unity_transformed}
\end{equation}
which, when substituted into the Lagrangian-based identity Eq.~\eqref{eq:delta_position}, yields Eq.~\eqref{eq:delta_from_j}.

This observation forms the conceptual basis of ULPT: since observational data are inherently obtained in Eulerian coordinates, it is natural to define fluctuations in Eulerian space and then transition to Lagrangian space for perturbative computations. 

Within this framework, the density contrast arises from two distinct physical effects:
\begin{itemize}
    \item \textbf{Jacobian deviation}:\\
        While the conventional definition of the density field emphasizes the inverse of the Jacobian determinant, $1/J(\boldsymbol{q})$, our formulation identifies the deviation from unity, defined as $\delta_{\rm J}(\boldsymbol{q}) \equiv 1 - J(\boldsymbol{q})$, as the physically meaningful quantity for describing density fluctuations. This quantity captures intrinsic density fluctuations and includes both linear and nonlinear contributions.

    \item \textbf{Displacement-mapping effect}:\\
    This effect accounts for the nonlinear remapping of the perturbed density field via the displacement vector $\boldsymbol{\Psi}$.
    It contributes only at nonlinear orders and modulates the spatial distribution of $\delta_{\rm J}$.
\end{itemize}

In Fourier space, Eq.~\eqref{eq:delta_fourier_from_j} expresses this decomposition as
\begin{align}
    \widetilde{\delta}(\boldsymbol{k}) &= \underbrace{\widetilde{\delta_{\rm J}}(\boldsymbol{k})}_{\text{Jacobian deviation}} \nonumber \\
&\quad + \underbrace{\int d^3q\, e^{-i\boldsymbol{k} \cdot \boldsymbol{q}}
\left[ e^{-i\boldsymbol{k} \cdot \boldsymbol{\Psi}(\boldsymbol{q})} - 1 \right] \delta_{\rm J}(\boldsymbol{q})}_{\text{Displacement-mapping effect}}.
\label{eq:delta_k_split}
\end{align}
where the first term represents the contribution from the Jacobian deviation, and the second term arises from the displacement-mapping effect.

By expanding the exponential in Eq.~\eqref{eq:delta_k_split} and performing the inverse Fourier transform, we obtain the corresponding expression in real space:
\begin{align}
    \delta(\boldsymbol{x}) &= \underbrace{\delta_{\rm J}(\boldsymbol{x})}_{\text{Jacobian deviation}} \nonumber \\
&\quad + \underbrace{\sum_{n=1}^\infty \frac{(-1)^n}{n!} 
    \partial_{i_1} \cdots \partial_{i_n}
    \left[ \Psi_{i_1}(\boldsymbol{x}) \cdots \Psi_{i_n}(\boldsymbol{x}) 
\delta_{\rm J}(\boldsymbol{x}) \right]}_{\text{Displacement-mapping effect}},
\label{eq:d_dJ_Psi}
\end{align}
where $\partial_i \equiv \partial / \partial x_i$. Note that all quantities in Eq.~\eqref{eq:d_dJ_Psi} are evaluated at the Eulerian coordinate $\boldsymbol{x}$.

Although the displacement-mapping term always appears in conjunction with $\delta_{\rm J}$ in the expression for the density contrast, their respective contributions to statistical observables, such as the power spectrum, can be systematically classified into two categories:  
(i) mixed terms involving both $\delta_{\rm J}$ and $\boldsymbol{\Psi}$, and  
(ii) pure displacement contributions originating solely from $\boldsymbol{\Psi}$.

A detailed discussion of how these contributions affect the power spectrum will be presented in Sec.~\ref{sec:power-spectrum}.

\section{Lagrangian Galaxy Bias}
\label{sec:GB}

\subsection{Standard Formulation of Lagrangian Bias}

To incorporate biased tracers such as galaxies into the Lagrangian framework, we begin by assuming the mass conservation relation~\cite{Matsubara:2008wx}:
\begin{equation}
    \bar{\rho}_{\mathrm{g}} [1 + \delta_{\mathrm{g}}(\boldsymbol{x})]\,d^3x = \bar{\rho}_{\mathrm{g}} [1 + \delta_{\mathrm{b}}(\boldsymbol{q})]\,d^3q.
\label{eq:galaxy_mass_conservation}
\end{equation}
Here, $\bar{\rho}_{\mathrm{g}}$ denotes the mean galaxy number density, $\delta_{\mathrm{g}}$ is the observed galaxy density contrast in Eulerian coordinates, and $\delta_{\mathrm{b}}$ represents the biased density contrast defined in Lagrangian coordinates. The subscripts ``g'' and ``b'' refer to ``galaxy'' and ``biased'', respectively.

Starting from the identity for the galaxy density field,
\begin{equation}
    \rho_{\mathrm{g}}(\boldsymbol{x}) = \int d^3x' \rho_{\mathrm{g}}(\boldsymbol{x}')\,\delta_{\mathrm{D}}(\boldsymbol{x} - \boldsymbol{x}'),
\end{equation}
and applying the Lagrangian-to-Eulerian mapping of Eq.~(\ref{eq:displacement}) along with the mass conservation law in Eq.~(\ref{eq:galaxy_mass_conservation}), we obtain the galaxy density contrast as
\begin{equation}
    1 + \delta_{\mathrm{g}}(\boldsymbol{x}) = \int d^3q\,[1 + \delta_{\mathrm{b}}(\boldsymbol{q})]\,\delta_{\mathrm{D}}(\boldsymbol{x} - \boldsymbol{q} - \boldsymbol{\Psi}(\boldsymbol{q})).
\label{eq:galaxy_density_real}
\end{equation}
Adopting the background condition
\begin{equation}
    1 = \int d^3q\,\delta_{\mathrm{D}}(\boldsymbol{x} - \boldsymbol{q}),
\end{equation}
and Fourier transforming, we obtain
\begin{equation}
    \widetilde{\delta}_{\mathrm{g}}(\boldsymbol{k}) = 
    \int d^3q\,e^{-i\boldsymbol{k}\cdot\boldsymbol{q}} \left\{e^{-i\boldsymbol{k}\cdot\boldsymbol{\Psi}(\boldsymbol{q})}[1 + \delta_{\mathrm{b}}(\boldsymbol{q})] - 1\right\}.
    \label{eq:delta_g_fourier}
\end{equation}

Furthermore, the biased density contrast can be exponentiated using the identity
\begin{equation}
\delta_{\mathrm{b}}(\boldsymbol{q}) = \left. \frac{d}{d\lambda} \exp[\lambda\,\delta_{\mathrm{b}}(\boldsymbol{q})] \right|_{\lambda=0},
\end{equation}
which allows for a unified exponential treatment. This representation has proven particularly useful in constructing LPT-based models due to its compact and mathematically tractable form~\cite{Matsubara:2008wx,Carlson:2012bu}.

\subsection{ULPT Formulation for Biased Tracers}

We now present the ULPT-based formulation for modeling galaxy bias, in which biased tracers are treated consistently within the unified Lagrangian framework. We begin with the identity for the galaxy density contrast,
\begin{equation}
    \delta_{\mathrm{g}}(\boldsymbol{x}) = \int d^3x'\,\delta_{\mathrm{g}}(\boldsymbol{x}')\,\delta_{\mathrm{D}}(\boldsymbol{x} - \boldsymbol{x}'),
\end{equation}
and rewrite the integrand in Lagrangian coordinates by applying the mapping $\boldsymbol{x}' = \boldsymbol{q} + \boldsymbol{\Psi}(\boldsymbol{q})$ and the volume element transformation $d^3x' = J(\boldsymbol{q})\,d^3q$. Using the mass conservation law modified for biased tracers in Eq.~\eqref{eq:galaxy_mass_conservation},
\begin{equation}
\delta_{\mathrm{g}}(\boldsymbol{q}) = \frac{1 + \delta_{\mathrm{b}}(\boldsymbol{q})}{J(\boldsymbol{q})} - 1,
\end{equation}
we arrive at
\begin{equation}
    \delta_{\mathrm{g}}(\boldsymbol{x}) = \int d^3q\,\left[\delta_{\rm J}(\boldsymbol{q}) + \delta_{\mathrm{b}}(\boldsymbol{q})\right]\,\delta_{\mathrm{D}}(\boldsymbol{x} - \boldsymbol{q} - \boldsymbol{\Psi}(\boldsymbol{q})),
\label{eq:galaxy_delta_real}
\end{equation}
which becomes in Fourier space:
\begin{equation}
    \widetilde{\delta}_{\mathrm{g}}(\boldsymbol{k}) = \int d^3q\,e^{-i\boldsymbol{k}\cdot\boldsymbol{q}}\,e^{-i\boldsymbol{k}\cdot\boldsymbol{\Psi}(\boldsymbol{q})}\left[\delta_{\rm J}(\boldsymbol{q}) + \delta_{\mathrm{b}}(\boldsymbol{q})\right].
\label{eq:galaxy_delta_fourier}
\end{equation}

In our formulation, the Lagrangian bias term $\delta_{\mathrm{b}}$ appears as a linear addition to the Jacobian deviation $\delta_{\rm J}$. As a result, the displacement-mapping effect, which is encoded in the exponential, modulates the combined contribution $[\delta_{\rm J} + \delta_{\mathrm{b}}]$ uniformly, without altering the bias structure itself. This ULPT formulation enables a consistent treatment of galaxy bias within the same theoretical framework developed for dark matter.

For instance, assuming a linear Lagrangian bias of the form $\delta_{\mathrm{b}} = b_1^{\mathrm{L}} \delta_{\rm J}$, where $b_1 = 1 + b_1^{\mathrm{L}}$ corresponds to the familiar Eulerian linear bias parameter, this linear bias contribution is fully compatible with the ULPT formulation and does not alter the structural separation between the Jacobian and the displacement-mapping effects.

While a detailed investigation of nonlinear bias effects is beyond the scope of this paper, we argue in Sec.~\ref{sec:galaxy_bias} that the nonlinear structure of the Jacobian deviation in the ULPT framework may provide new insights into the parameterization of nonlinear galaxy bias.

\section{REAL-TO-REDSHIFT SPACE MAPPING}
\label{sec:RRM}

\subsection{Standard Formulation of RSD}

\subsubsection{Density-Based Representation in the Eulerian Picture}

The mapping from real-space Eulerian coordinates $\boldsymbol{x}$ to observed redshift-space coordinates $\boldsymbol{s}$ is given by
\begin{equation}
\boldsymbol{s} = \boldsymbol{x} + \frac{\boldsymbol{v}(\boldsymbol{x}) \cdot \hat{\boldsymbol{n}}}{H} \, \hat{\boldsymbol{n}}, 
\label{eq:rsd_mapping}
\end{equation}
where $\hat{\boldsymbol{n}}$ denotes the unit vector along the line of sight, $\boldsymbol{v}(\boldsymbol{x})$ is the peculiar velocity field in comoving coordinates, and $H$ is the Hubble parameter.

Assuming number conservation, the galaxy number densities in real and redshift space are related by
\begin{equation}
    \rho_{\rm gs}(\boldsymbol{s},\hat{\boldsymbol{n}}) \, d^3 s
    = \rho_{\mathrm g}(\boldsymbol{x}) \, d^3 x,
\label{eq:number_conservation}
\end{equation}
where the subscript ``s'' indicates redshift-space quantities; thus, $\rho_{\rm gs}$ represents the galaxy number density field in redshift space. 

Quantities defined in redshift space generally depend on the line-of-sight direction \(\hat{\boldsymbol{n}}\). In what follows, however, we omit the explicit dependence on \(\hat{\boldsymbol{n}}\) for notational simplicity. For example, we write \(\rho_{\mathrm{gs}}(\boldsymbol{s})\) instead of \(\rho_{\mathrm{gs}}(\boldsymbol{s}, \hat{\boldsymbol{n}})\). Throughout this paper, we assume that all quantities labeled with the subscript ``s'', indicating redshift-space quantities, implicitly depend on \(\hat{\boldsymbol{n}}\).

To express the redshift-space density field in terms of real-space quantities, we begin with the identity
\begin{equation}
    \rho_{\mathrm{gs}}(\boldsymbol{s}) = \int d^3 s' \, \rho_{\rm gs}(\boldsymbol{s}') \, 
    \delta_{\rm D}(\boldsymbol{s} - \boldsymbol{s}'), 
    \label{eq:delta_identity_rsd}
\end{equation}
and apply the change of variables using Eq.~\eqref{eq:rsd_mapping}, yielding
\begin{equation}
    \rho_{\rm gs}(\boldsymbol{s}) = \int d^3 x \, \rho_{\rm g}(\boldsymbol{x}) \, \delta_{\rm D}\left(\boldsymbol{s} - \boldsymbol{x} - \frac{\boldsymbol{v}(\boldsymbol{x}) \cdot \hat{\boldsymbol{n}}}{H} \, \hat{\boldsymbol{n}}\right). 
\label{eq:rsd_density}
\end{equation}
Accordingly, the redshift-space galaxy density contrast is written as
\begin{eqnarray}
    && 1 + \delta_{\rm gs}(\boldsymbol{s}) \nonumber \\
    &=& \int d^3x \, [1 + \delta_{\rm g}(\boldsymbol{x})] \, \delta_{\rm D}\left(\boldsymbol{s} - \boldsymbol{x} - \frac{\boldsymbol{v}(\boldsymbol{x}) \cdot \hat{\boldsymbol{n}}}{H} \, \hat{\boldsymbol{n}}\right). 
\label{eq:rsd_contrast}
\end{eqnarray}
The term $1$ on the left-hand side of Eq.~\eqref{eq:rsd_contrast} corresponds to the contribution from the background galaxy number density. To confirm this interpretation, we consider the case where both the density and velocity fields vanish. In this case, Eq.~\eqref{eq:rsd_contrast} reduces to
\begin{equation}
    1 = \int d^3 x \, \delta_{\rm D}(\boldsymbol{s} - \boldsymbol{x}), 
    \label{eq:background_identity}
\end{equation}
which recovers the background value as expected.

Substituting Eq.~\eqref{eq:background_identity} into Eq.~\eqref{eq:rsd_contrast}, we obtain the fluctuation component of the redshift-space galaxy density:
\begin{eqnarray}
    \delta_{\rm gs}(\boldsymbol{s})
    &=&
    \int d^3 x \, \Bigg\{[1 + \delta_{\rm g}(\boldsymbol{x})] \, \delta_{\rm D}\left(\boldsymbol{s} - \boldsymbol{x} - \frac{\boldsymbol{v}(\boldsymbol{x}) \cdot \hat{\boldsymbol{n}}}{H} \, \hat{\boldsymbol{n}}\right) \nonumber \\
    &&\hspace{1.5cm}- \delta_{\rm D}(\boldsymbol{s} - \boldsymbol{x}) \Bigg\}. 
\label{eq:rsd_fluctuation}
\end{eqnarray}
Taking the Fourier transform of Eq.~\eqref{eq:rsd_fluctuation}, we obtain
\begin{equation}
    \widetilde{\delta}_{\rm gs}(\boldsymbol{k}) = \int d^3 x \, e^{-i \boldsymbol{k} \cdot \boldsymbol{x}} \left\{ e^{-i \boldsymbol{k} \cdot \hat{\boldsymbol{n}} \frac{\boldsymbol{v}(\boldsymbol{x}) \cdot \hat{\boldsymbol{n}}}{H}} [1 + \delta_{\rm g}(\boldsymbol{x})] - 1 \right\}. 
    \label{eq:rsd_fourier}
\end{equation}
A representative example of an RSD model that takes this expression as a starting point is the distribution function approach~\cite{Seljak:2011tx}.

\subsubsection{Jacobian-Based Representation in the Eulerian Picture}
\label{sec:rsd-euler}

An alternative formulation in the Eulerian picture makes explicit use of the Jacobian determinant associated with the transformation from real-space to redshift-space coordinates.

The Jacobian determinant for the real-to-redshift space mapping is defined as
\begin{equation}
    J_{\rm s}(\boldsymbol{x}) \equiv \det\left( \frac{\partial \boldsymbol{s}}{\partial \boldsymbol{x}} \right), \label{eq:js_def}
\end{equation}
which implies the volume element transforms as
\begin{equation}
    d^3 s = J_{\rm s}(\boldsymbol{x}) \, d^3 x. 
    \label{eq:js_volume}
\end{equation}

This Jacobian determinant can be explicitly evaluated. Assuming the line-of-sight direction is aligned with the $z$-axis, i.e., $\hat{\boldsymbol{n}} = \hat{\boldsymbol{z}}$, we obtain
\begin{equation}
    J_{\rm s}(\boldsymbol{x}) = 1 + \frac{1}{H} \frac{\partial v_z(\boldsymbol{x})}{\partial z}, \label{eq:js_eval}
\end{equation}
where $v_z$ is the $z$-component of the peculiar velocity field.

To isolate the contribution from the background density in redshift space, we write
\begin{equation}
    1 = \int d^3  s' \, \delta_{\rm D}(\boldsymbol{s} - \boldsymbol{s}'),
    \label{eq:rsd_background}
\end{equation}
and substitute the mapping \eqref{eq:rsd_mapping} and volume transformation \eqref{eq:js_volume} into Eq.~\eqref{eq:rsd_background}. Using Eq.~\eqref{eq:js_eval}, we obtain
\begin{equation}
    1 = \int d^3 x \left( 1 + \frac{1}{H} \frac{\partial v_z(\boldsymbol{x})}{\partial z} \right) \delta_{\rm D}\left( \boldsymbol{s} - \boldsymbol{x} - \frac{v_z(\boldsymbol{x})}{H} \hat{\boldsymbol{z}} \right). \label{eq:rsd_background_transformed}
\end{equation}

Inserting this expression into the redshift-space contrast formula in Eq.~\eqref{eq:rsd_contrast}, we arrive at
\begin{eqnarray}
    \delta_{\rm gs}(\boldsymbol{s}) &=& \int d^3 x\, \left[ \delta_{\rm g}(\boldsymbol{x}) - \frac{1}{H} \frac{\partial v_z(\boldsymbol{x})}{\partial z} \right] \nonumber\\
    &\times&
    \delta_{\rm D}\left( \boldsymbol{s} - \boldsymbol{x} - \frac{v_z(\boldsymbol{x})}{H} \hat{\boldsymbol{z}} \right). \label{eq:rsd_contrast_jacobian}
\end{eqnarray}

Taking the Fourier transform of Eq.~\eqref{eq:rsd_contrast_jacobian}, we obtain
\begin{equation}
    \widetilde{\delta}_{\rm gs}(\boldsymbol{k}) = \int d^3 x \, e^{-i \boldsymbol{k} \cdot \boldsymbol{x}} \, e^{-i k_z \frac{v_z(\boldsymbol{x})}{H}} \left[ \delta_{\rm g}(\boldsymbol{x}) - \frac{1}{H} \frac{\partial v_z(\boldsymbol{x})}{\partial z} \right], \label{eq:rsd_fourier_jacobian}
\end{equation}
where $k_z \equiv \boldsymbol{k} \cdot \hat{\boldsymbol{z}}$.

The integrand in Eq.~\eqref{eq:rsd_fourier_jacobian} contains the leading-order linear redshift-space distortion (RSD) effect, known as the Kaiser effect~\cite{Kaiser:1987qv}, through the combination $[\delta_{\rm g} - (1/H)\, \partial_z v_z]$. Moreover, it includes nonlinear contributions arising from both the density and velocity fields, as well as the exponential factor that encodes nonlinear velocity effects beyond linear theory.

A representative model based on this expression is the one developed by Taruya, Nishimichi, and Saito~\cite{Taruya:2010mx}, commonly referred to as the TNS model. This model has been successfully applied to galaxy clustering data from the 
the Baryon Oscillation Spectroscopic Survey (BOSS) survey and beyond~\cite{BOSS:2016psr}, offering an accurate description of RSD effects over a wide range of scales.

\subsubsection{Density-Based Representation in the Lagrangian Picture}

The transformation from Lagrangian coordinates $\boldsymbol{q}$ to redshift-space coordinates $\boldsymbol{s}$ is given by
\begin{equation}
    \boldsymbol{s} = \boldsymbol{q} + \boldsymbol{\Psi}(\boldsymbol{q}) + \frac{\dot{\boldsymbol{\Psi}}(\boldsymbol{q}) \cdot \hat{\boldsymbol{n}}}{H} \hat{\boldsymbol{n}},
    \label{eq:s_q_mapping}
\end{equation}
where we have used the fact that the displacement field $\boldsymbol{\Psi}$ is related to the peculiar velocity via $\dot{\boldsymbol{\Psi}} \equiv d\boldsymbol{\Psi}/dt=\boldsymbol{v}(\boldsymbol{q} + \boldsymbol{\Psi}(\boldsymbol{q}))$.

For notational simplicity, we define the redshift-space displacement vector as
\begin{equation}
    \boldsymbol{\Psi}_{\rm s}(\boldsymbol{q}) \equiv \boldsymbol{\Psi}(\boldsymbol{q}) + \frac{\dot{\boldsymbol{\Psi}}(\boldsymbol{q}) \cdot \hat{\boldsymbol{n}}}{H} \hat{\boldsymbol{n}}. \label{eq:rs_displacement}
\end{equation}

Substituting Eq.~\eqref{eq:s_q_mapping} and the bias relation in Eq.~\eqref{eq:galaxy_density_real} into Eq.~\eqref{eq:rsd_contrast}, we obtain the redshift-space galaxy density contrast:
\begin{equation}
    1 + \delta_{\mathrm{gs}}(\boldsymbol{s}) = \int d^3 q \, [1 + \delta_{\rm b}(\boldsymbol{q})] \, \delta_{\rm D}(\boldsymbol{s} - \boldsymbol{q} - \boldsymbol{\Psi}_{\rm s}(\boldsymbol{q})). 
    \label{eq:delta_gs_real}
\end{equation}

Adopting the following expression for the background contribution
\begin{eqnarray}
    1 = \int d^3q\, \delta_{\rm D}(\boldsymbol{s} - \boldsymbol{q}),
\end{eqnarray}
and taking the Fourier transform, Eq.~\eqref{eq:delta_gs_real} becomes
\begin{equation}
    \widetilde{\delta}_{\mathrm{gs}}(\boldsymbol{k}) = \int d^3 q \, 
    e^{-i \boldsymbol{k} \cdot \boldsymbol{q}} \, 
    \Big\{e^{-i \boldsymbol{k} \cdot \boldsymbol{\Psi}_{\rm s}(\boldsymbol{q})} 
    \left[1 + \delta_{\rm b}(\boldsymbol{q}) \right] - 1\Big\}. 
    \label{eq:delta_gs_fourier}
\end{equation}

This expression is structurally analogous to Eq.~\eqref{eq:delta_g_fourier}, meaning that the displacement vector appears exclusively in the exponential, which makes it particularly amenable to treatment within a perturbative expansion. The key difference is that the displacement vector $\boldsymbol{\Psi}_{\rm s}$ now includes RSD effects.

\subsection{ULPT Formulation in Redshift Space}
\label{sec:core_rsd}

In this section, we present the ULPT-based formulation of RSD, which extends our unified treatment of the density field to redshift space. This formulation is analogous in structure to the Jacobian-based approach previously introduced in the Eulerian picture (see Sec.~\ref{sec:rsd-euler}), but is grounded in the ULPT framework that explicitly separates the Jacobian deviation from the displacement-mapping effect.

The Jacobian determinant associated with the mapping from Lagrangian coordinates $\boldsymbol{q}$ to redshift-space coordinates $\boldsymbol{s}$ is defined as
\begin{equation}
    J_{\rm s}^{\mathrm{L}}(\boldsymbol{q}) \equiv \det \left( \frac{\partial \boldsymbol{s}}{\partial \boldsymbol{q}} \right), \label{eq:jacobian_lagrangian}
\end{equation}
where the superscript ``L'' indicates that the determinant is evaluated in the Lagrangian frame. This implies the volume element transforms as
\begin{equation}
    d^3 s = J_{\rm s}^{\mathrm{L}}(\boldsymbol{q}) \, d^3 q. 
    \label{eq:volume_lagrangian}
\end{equation}

The mass conservation relation for galaxies under the mapping from Lagrangian to redshift-space coordinates reads
\begin{equation}
    [1 + \delta_{\mathrm{gs}}(\boldsymbol{s})]\,d^3s = 
    [1 + \delta_{\mathrm{b}}(\boldsymbol{q})]\,d^3q,
\end{equation}
which leads to
\begin{equation}
    \delta_{\mathrm{gs}}(\boldsymbol{s}) = \frac{1 + \delta_{\mathrm{b}}(\boldsymbol{q})}{J^{\mathrm{L}}_{\rm s}(\boldsymbol{q})} - 1.
\label{eq:dgs_J}
\end{equation}

We start from the identity for the galaxy density contrast in redshift space,
\begin{equation}
    \delta_{\mathrm{gs}}(\boldsymbol{s}) = \int d^3s'\,\delta_{\mathrm{gs}}(\boldsymbol{s}')\,\delta_{\mathrm{D}}(\boldsymbol{s} - \boldsymbol{s}'),
\end{equation}
and rewrite the integrand in Lagrangian coordinates. This yields
\begin{equation}
    \delta_{\mathrm{gs}}(\boldsymbol{s}) = \int d^3q\,\left[\delta_{\rm Js}(\boldsymbol{q}) + \delta_{\mathrm{b}}(\boldsymbol{q})\right]\,\delta_{\mathrm{D}}(\boldsymbol{s} - \boldsymbol{q} - \boldsymbol{\Psi}_{\rm s}(\boldsymbol{q})),
    \label{eq:delta_gs_realspace_lpt}
\end{equation}
where 
\begin{align}
    \delta_{\rm Js}(\boldsymbol{q}) + \delta_{\mathrm{b}}(\boldsymbol{q})
    &= J^{\mathrm{L}}_{\mathrm{s}}(\boldsymbol{q}) \, \delta_{\mathrm{gs}}(\boldsymbol{q} + \boldsymbol{\Psi}_{\mathrm{s}}(\boldsymbol{q})) \nonumber \\
    &= J^{\mathrm{L}}_{\mathrm{s}}(\boldsymbol{q}) \left[ \frac{1 + \delta_{\mathrm{b}}(\boldsymbol{q})}{J^{\mathrm{L}}_{\mathrm{s}}(\boldsymbol{q})}  - 1 \right]\nonumber \\
    &= \left[ 1 - J^{\mathrm{L}}_{\mathrm{s}}(\boldsymbol{q}) \right] + \delta_{\mathrm{b}}(\boldsymbol{q}),
    \label{eq:delta_Js_decomposition}
\end{align}
and we define the Jacobian deviation in redshift space as
\begin{equation}
    \delta_{\rm Js}(\boldsymbol{q}) \equiv 1 - J^{\mathrm{L}}_{\rm s}(\boldsymbol{q}).
\end{equation}

Assuming the line-of-sight direction $\hat{\boldsymbol{n}}$ is aligned with the $z$-axis, the redshift-space Jacobian deviation can be written explicitly as
\begin{align}
\delta_{\rm Js}(\boldsymbol{q}) =\
& - \Psi_{i,i}(\boldsymbol{q})
  - \frac{1}{H} \dot{\Psi}_{z,z}(\boldsymbol{q}) \nonumber \\
& - \frac{1}{2} \left[
    \Psi_{i,i} \Psi_{j,j}
    - \Psi_{i,j} \Psi_{j,i}
  \right] \nonumber \\
& - \frac{1}{2} \left[
    \frac{2}{H} \dot{\Psi}_{z,z} \Psi_{i,i}
    - \frac{2}{H} \Psi_{i,z} \dot{\Psi}_{z,i}
      \right] \nonumber \\
& - \frac{1}{6} \epsilon_{ijk} \epsilon_{lmn}
    \Psi_{i,l} \Psi_{j,m} \Psi_{k,n} \nonumber \\
& - \frac{1}{6} \left[
    \frac{1}{H} \epsilon_{ijz} \epsilon_{lmn}
    \Psi_{i,l} \Psi_{j,m} \dot{\Psi}_{z,n}
    + (i,j \leftrightarrow z)
\right].
\label{eq:deltaJs}
\end{align}

Taking the Fourier transform of the redshift-space galaxy density contrast, we obtain
\begin{equation}
    \widetilde{\delta}_{\mathrm{gs}}(\boldsymbol{k}) = \int d^3q\,e^{-i\boldsymbol{k} \cdot \boldsymbol{q}}\,e^{-i\boldsymbol{k} \cdot \boldsymbol{\Psi}_{\rm s}(\boldsymbol{q})}\left[\delta_{\rm Js}(\boldsymbol{q}) + \delta_{\mathrm{b}}(\boldsymbol{q})\right].
 \label{eq:delta_gs_fourier_lpt}
\end{equation}

This expression retains the same structural form as its real-space counterpart in Eq.~\eqref{eq:galaxy_delta_fourier}, but now interpreted within the ULPT framework extended to redshift space. In this formulation, both the displacement vector and the Jacobian deviation are modified to include velocity contributions along the line of sight. The ULPT structure, based on the combination of the Jacobian deviation and the displacement-mapping effect, remains intact, allowing RSD to be described consistently within the same theoretical framework developed for real-space and biased tracers. In the integrand of Eq.~\eqref{eq:delta_gs_fourier_lpt}, the term $[\delta_{\rm Js} + \delta_{\rm b}]$ includes the linear contributions associated with the well-known Kaiser effect, while also containing nonlinear corrections beyond linear theory.

\section{Pre-to-Post-Reconstruction Mapping}
\label{sec:PRM}

\subsection{Standard Formulation of Reconstruction}

To reconstruct the large-scale distribution of galaxies, we adopt the simplest reconstruction algorithm originally proposed in Ref.~\cite{Eisenstein:2006nk}. In this method, the displacement field used in reconstruction is estimated from the observed galaxy density fluctuations in redshift space as follows:
\begin{equation}
    \boldsymbol{t}_{\mathrm{gs}}(\boldsymbol{s}) = i \int \frac{d^3 k}{(2\pi)^3} \, e^{i \boldsymbol{k} \cdot \boldsymbol{s}} \, \boldsymbol{R}(\boldsymbol{k}) \, \widetilde{\delta}_{\mathrm{gs}}(\boldsymbol{k}),
    \label{eq:tgs}
\end{equation}
where the kernel $\boldsymbol{R}(\boldsymbol{k})$ is given by
\begin{equation}
    \boldsymbol{R}(\boldsymbol{k}) = \frac{\boldsymbol{k}}{k^2} \, 
    \left( -\frac{W_{\rm G}(k R)}{b_{1,\mathrm{fid}}} \right),
\end{equation}
with $b_{1,\mathrm{fid}}$ denoting the fiducial linear galaxy bias used in the reconstruction procedure, and $W_{\rm G}(k R) = \exp(-k^2 R^2 / 2)$ being a Gaussian smoothing filter with scale $R$.

As a reminder, the subscript ``gs'' in $\boldsymbol{t}_{\mathrm{gs}}$ and $\delta_{\mathrm{gs}}$ indicates that these quantities are constructed from observed spectroscopic galaxy samples. Therefore, they implicitly incorporate both galaxy bias and RSD. In particular, because of the anisotropic nature of RSD, both $\boldsymbol{t}_{\mathrm{gs}}$ and $\delta_{\mathrm{gs}}$ depend on the line-of-sight direction $\hat{\boldsymbol{n}}$.

The coordinate transformation from redshift space to the reconstructed space is then defined as
\begin{equation}
    \boldsymbol{x}_{\mathrm{rec}} = \boldsymbol{s} + \boldsymbol{t}_{\mathrm{gs}}(\boldsymbol{s}),
    \label{eq:mapping_rec}
\end{equation}
where the subscript ``rec'' denotes ``reconstructed.''

Assuming number conservation, the galaxy number density before and after reconstruction satisfies
\begin{equation}
    \rho_{\mathrm{gs,rec}}(\boldsymbol{x}_{\mathrm{rec}}) \, d^3 x_{\mathrm{rec}} 
    = \rho_{\mathrm{gs}}(\boldsymbol{s}) \, d^3 s.
    \label{eq:cons_rec}
\end{equation}

\subsubsection{Standard Eulerian Representation}

To express the reconstructed density field in terms of the original field, we start from the identity
\begin{equation}
    \rho_{\mathrm{gs,rec}}(\boldsymbol{x}_{\mathrm{rec}}) 
    = \int d^3 x'_{\mathrm{rec}} \, \rho_{\mathrm{gs,rec}}(\boldsymbol{x}'_{\mathrm{rec}}) \, \delta_{\rm D}(\boldsymbol{x}_{\mathrm{rec}} - \boldsymbol{x}'_{\mathrm{rec}}).
\end{equation}
Substituting the coordinate mapping in Eq.~\eqref{eq:mapping_rec} and the number conservation relation in Eq.~\eqref{eq:cons_rec} into this expression yields
\begin{equation}
    \rho_{\mathrm{gs,rec}}(\boldsymbol{x}_{\mathrm{rec}}) 
    = \int d^3s \, \rho_{\mathrm{gs}}(\boldsymbol{s}) \, 
    \delta_{\rm D}(\boldsymbol{x}_{\mathrm{rec}} - \boldsymbol{s} - \boldsymbol{t}_{\mathrm{gs}}(\boldsymbol{s})).
    \label{eq:rho_gsrec}
\end{equation}

A key feature of the reconstruction procedure is that the same displacement $\boldsymbol{t}_{\mathrm{gs}}$ is applied not only to the galaxy sample but also to a synthetic random catalog. Denoting the number density of the random particles as $\rho_{\mathrm{ran}}$, the number conservation condition implies
\begin{equation}
    \rho_{\mathrm{ran,rec}}(\boldsymbol{x}_{\mathrm{rec}}) \, d^3 x_{\mathrm{rec}}
    = \rho_{\mathrm{ran}}(\boldsymbol{s}) \, d^3 s,
\end{equation}
which leads to
\begin{equation}
    \rho_{\mathrm{ran,rec}}(\boldsymbol{x}_{\mathrm{rec}}) 
    = \int d^3 s \, \rho_{\mathrm{ran}}(\boldsymbol{s}) \, 
    \delta_{\rm D}(\boldsymbol{x}_{\mathrm{rec}} - \boldsymbol{s} - \boldsymbol{t}_{\mathrm{gs}}(\boldsymbol{s})).
    \label{eq:ran_rec}
\end{equation}

Assuming that the random field is uniform with the mean galaxy density $\bar{\rho}_{\rm g}$, we have
\begin{equation}
    \rho_{\mathrm{ran}}(\boldsymbol{s}) 
    = \bar{\rho}_{\rm g} \int d^3 s' \, \delta_{\rm D}(\boldsymbol{s} - \boldsymbol{s}'),
\end{equation}
and hence,
\begin{equation}
    \rho_{\mathrm{ran,rec}}(\boldsymbol{x}_{\mathrm{rec}}) 
    = \bar{\rho}_{\rm g} \int d^3 s \, \delta_{\rm D}(\boldsymbol{x}_{\mathrm{rec}} - \boldsymbol{s} - \boldsymbol{t}_{\mathrm{gs}}(\boldsymbol{s})).
\end{equation}

The reconstructed galaxy density contrast is then defined by subtracting the reconstructed random field:
\begin{equation}
    \delta_{\mathrm{gs,rec}}(\boldsymbol{x}_{\mathrm{rec}}) = \frac{\rho_{\mathrm{gs,rec}}(\boldsymbol{x}_{\mathrm{rec}}) - \rho_{\mathrm{ran,rec}}(\boldsymbol{x}_{\mathrm{rec}})}{\bar{\rho}_{\rm g}},
    \label{eq:delta_rec}
\end{equation}
which yields~\cite{Sugiyama:2020uil,Shirasaki:2020vkk}
\begin{equation}
    \delta_{\mathrm{gs,rec}}(\boldsymbol{x}_{\mathrm{rec}}) 
    = \int d^3 s \, \delta_{\mathrm{gs}}(\boldsymbol{s}) \, \delta_{\rm D}(\boldsymbol{x}_{\mathrm{rec}} - \boldsymbol{s} - \boldsymbol{t}_{\mathrm{gs}}(\boldsymbol{s})).
    \label{eq:delta_rec_realspace}
\end{equation}

In Fourier space, this becomes
\begin{equation}
    \widetilde{\delta}_{\mathrm{gs,rec}}(\boldsymbol{k}) 
    = \int d^3 s \, e^{-i \boldsymbol{k} \cdot \boldsymbol{s}} \, e^{-i \boldsymbol{k} \cdot \boldsymbol{t}_{\mathrm{gs}}(\boldsymbol{s})} \, \delta_{\mathrm{gs}}(\boldsymbol{s}).
    \label{eq:delta_rec_fourier}
\end{equation}

Equation~\eqref{eq:delta_rec_fourier} highlights a central aspect of the reconstruction framework: in Fourier space, the displacement vector $\boldsymbol{t}_{\rm gs}$, computed from the pre-reconstruction density field, appears in the exponent and multiplies the pre-reconstruction density fluctuation. Within this formulation, the effect of reconstruction arises solely from nonlinear contributions, making it clear that the density fluctuations remain unchanged at the linear level. Furthermore, starting from this expression, it has been demonstrated that the nonlinear damping of the BAO signal in the post-reconstruction power spectrum can be accurately modeled using a single Gaussian-form exponential damping factor, just as in the pre-reconstruction case~\cite{Sugiyama:2024eye,Chen:2024tfp}.

\subsubsection{Standard Lagrangian Reconstruction}

The transformation from Lagrangian coordinates to the post-reconstruction Eulerian coordinates can be derived by substituting Eq.~\eqref{eq:s_q_mapping} into Eq.~\eqref{eq:mapping_rec}, leading to
\begin{equation}
    \boldsymbol{x}_{\mathrm{rec}} = \boldsymbol{q} + \boldsymbol{\Psi}_{s}(\boldsymbol{q}) + \boldsymbol{t}_{\mathrm{gs}}(\boldsymbol{q} + \boldsymbol{\Psi}_{s}(\boldsymbol{q})).
    \label{eq:x_rec}
\end{equation}
For notational simplicity, we define the post-reconstruction displacement vector as
\begin{equation}
    \boldsymbol{\Psi}_{\rm s,\mathrm{rec}}(\boldsymbol{q}) \equiv \boldsymbol{\Psi}_{\rm s}(\boldsymbol{q}) + \boldsymbol{t}_{\mathrm{gs}}(\boldsymbol{q} + \boldsymbol{\Psi}_{\rm s}(\boldsymbol{q})).
    \label{eq:psi_s_rec}
\end{equation}

Substituting the Lagrangian-space expression for the redshift-space galaxy density field given by Eq.~\eqref{eq:delta_gs_real} into Eq.~\eqref{eq:rho_gsrec}, we obtain the post-reconstruction density field as
\begin{eqnarray}
    && \rho_{\mathrm{gs,rec}}(\boldsymbol{x}_{\mathrm{rec}}) \nonumber \\
    &=& \bar{\rho}_{\rm g} \int d^3 q \, [1 + \delta_{\rm b}(\boldsymbol{q})] \, \delta_{\rm D}(\boldsymbol{x}_{\mathrm{rec}} - \boldsymbol{q} - \boldsymbol{\Psi}_{\rm s,\mathrm{rec}}(\boldsymbol{q})).
    \label{eq:rho_gs_rec_lag}
\end{eqnarray}

For the random density field, we adopt the following representation:
\begin{equation}
    \rho_{\mathrm{ran}}(\boldsymbol{s}) 
    = \bar{\rho}_{\rm g} \int d^3 q \, \delta_{\rm D}(\boldsymbol{s} - \boldsymbol{q}).
    \label{eq:rho_ran_lag}
\end{equation}
Applying the same mapping, the corresponding post-reconstruction random field becomes
\begin{equation}
    \rho_{\mathrm{ran,rec}}(\boldsymbol{x}_{\mathrm{rec}}) 
    = \bar{\rho}_{\rm g} \int d^3 q \, \delta_{\rm D}(\boldsymbol{x}_{\mathrm{rec}} - \boldsymbol{q} - \boldsymbol{t}_{\mathrm{gs}}(\boldsymbol{q})).
\end{equation}

Combining the two contributions, the reconstructed galaxy density contrast is given by
\begin{align}
    \delta_{\mathrm{gs,rec}}(\boldsymbol{x}_{\mathrm{rec}}) 
    = & \int d^3 q \, [1 + \delta_{\rm b}(\boldsymbol{q})] \, \delta_{\rm D}(\boldsymbol{x}_{\mathrm{rec}} - \boldsymbol{q} - \boldsymbol{\Psi}_{\rm s,\mathrm{rec}}(\boldsymbol{q})) \nonumber \\
    &- \int d^3 q \, \delta_{\rm D}(\boldsymbol{x}_{\mathrm{rec}} - \boldsymbol{q} - \boldsymbol{t}_{\mathrm{gs}}(\boldsymbol{q})).
\end{align}

In Fourier space, this becomes
\begin{eqnarray}
    \widetilde{\delta}_{\mathrm{gs,rec}}(\boldsymbol{k}) 
    &=& \int d^3q \, e^{-i \boldsymbol{k} \cdot \boldsymbol{q}} \nonumber \\
    &\times& \left\{ e^{-i \boldsymbol{k} \cdot \boldsymbol{\Psi}_{\rm s,\mathrm{rec}}(\boldsymbol{q})} [1 + \delta_{\rm b}(\boldsymbol{q})] - e^{-i \boldsymbol{k} \cdot \boldsymbol{t}_{\mathrm{gs}}(\boldsymbol{q})} \right\}.
    \label{eq:delta_rec_fourier_lag}
\end{eqnarray}

In this expression, if both the redshift-space displacement vector \( \boldsymbol{\Psi}_{\rm s,\mathrm{rec}} \) and the reconstruction displacement field \( \boldsymbol{t}_{\mathrm{gs}} \) are linearly approximated, the result corresponds to the so-called post-reconstruction ZA. The limitations of this approximation have been thoroughly investigated by Ref.~\cite{Sugiyama:2024eye}.

One well-known issue is that this approach predicts two distinct Gaussian damping factors in the BAO feature of the power spectrum~\cite{Padmanabhan:2008dd,Seo:2015eyw,White:2015eaa,Chen:2019lpf}, in contrast to the single Gaussian damping factor expected from a consistent treatment. The origin of this discrepancy lies in the linear treatment of \( \boldsymbol{t}_{\mathrm{gs}}(\boldsymbol{q}) \) in Eq.~\eqref{eq:delta_rec_fourier_lag}. Since \( \boldsymbol{t}_{\mathrm{gs}} \) is constructed from the observed galaxy density contrast, it implicitly contains nonperturbative IR effects that arise from long-wavelength displacements. When this field is linearized, these IR effects are removed, leading to an incorrect representation of the nonlinear behavior of the BAO signal.

Moreover, linearizing \( \boldsymbol{t}_{\mathrm{gs}}(\boldsymbol{q}) \) destroys the proper cancellation of the IR effect expected in the IR limit, resulting in the emergence of the two separate damping factors.

In contrast, Eq.~\eqref{eq:delta_rec_fourier} avoids this problem. The key difference is that \( \boldsymbol{t}_{\mathrm{gs}} \) is evaluated at Eulerian positions \( \boldsymbol{s} \), and thus effectively retains the full nonlinear structure \( \boldsymbol{t}_{\mathrm{gs}}(\boldsymbol{q} + \boldsymbol{\Psi}_{\rm s}(\boldsymbol{q})) \). In this form, the IR-sensitive part of \( \boldsymbol{t}_{\mathrm{gs}} \) cancels internally and no longer contributes to the damping, even if a linear approximation is applied to the remaining structure. As a result, the IR cancellation is preserved and a single Gaussian damping factor correctly describes the reconstructed BAO signal.

Furthermore, Ref.~\cite{Sugiyama:2024eye} shows that even when starting from Eq.~\eqref{eq:delta_rec_fourier_lag}, the correct single Gaussian damping factor can be recovered if the nonlinear IR-sensitive structure of \( \boldsymbol{t}_{\mathrm{gs}}(\boldsymbol{q}) \) is treated properly.

\subsection{ULPT Formulation of Reconstruction}

The reconstructed density fluctuation has already been expressed in Eq.~\eqref{eq:delta_rec_realspace} in terms of the observed galaxy density contrast in redshift space, $\delta_{\mathrm{gs}}(\mathbf{s})$, and the reconstruction displacement vector, $\mathbf{t}_{\mathrm{gs}}(\mathbf{s})$. We now reinterpret this expression within the ULPT framework, which provides a unified Lagrangian description of the reconstruction process.

To develop a Lagrangian formulation of reconstruction, we substitute into Eq.~\eqref{eq:delta_rec_realspace} the mapping from Lagrangian to redshift-space coordinates, $\boldsymbol{s} = \boldsymbol{q} + \boldsymbol{\Psi}_{\rm s}(\boldsymbol{q})$, together with the associated volume element transformation, $d^3 s = J_{\rm s}^{\rm L}(\boldsymbol{q})\, d^3 q$. Using Eqs.~\eqref{eq:delta_Js_decomposition} and~\eqref{eq:psi_s_rec}, the reconstructed galaxy density contrast in configuration space becomes
\begin{align}
    \delta_{\rm gs,rec}(\boldsymbol{x}_{\rm rec})
    = \int d^3 q\, &\left[ \delta_{\rm Js}(\boldsymbol{q}) + \delta_{\rm b}(\boldsymbol{q}) \right] \notag \\
    &\times \delta_{\rm D} \left( \boldsymbol{x}_{\rm rec} - \boldsymbol{q} - \boldsymbol{\Psi}_{\rm s,rec}(\boldsymbol{q}) \right),
    \label{eq:delta_rec_config}
\end{align}
which, upon Fourier transformation, yields
\begin{equation}
    \widetilde{\delta}_{\rm gs,rec}(\boldsymbol{k}) 
    = \int d^3 q\, e^{-i \boldsymbol{k} \cdot \boldsymbol{q}}\, e^{-i \boldsymbol{k} \cdot \boldsymbol{\Psi}_{\rm s,rec}(\boldsymbol{q})} \left[ \delta_{\rm Js}(\boldsymbol{q}) + \delta_{\rm b}(\boldsymbol{q}) \right].
    \label{eq:delta_rec_fourier_lag_J}
\end{equation}

This ULPT-based formulation provides a consistent description that encompasses all the previously discussed cases. Specifically, it includes dark matter in real space (Eq.~\eqref{eq:delta_from_j}), biased tracers (Eq.~\eqref{eq:galaxy_delta_real}), and redshift-space distortions (Eq.~\eqref{eq:delta_gs_realspace_lpt}) as special instances. In each of these cases, the density fluctuation is decomposed into two conceptually distinct components: the \emph{Jacobian deviation}, which generates intrinsic number density fluctuations, and the \emph{displacement-mapping effect}, which modulates their spatial distribution through coordinate transformation.

The differences among the cases arise solely from the specific physical ingredients represented in each term:
\begin{itemize}
    \item whether RSDs are incorporated into the Jacobian deviation,
    \item whether the biased fluctuation contributes additively to the Jacobian deviation,
    \item and whether the displacement vector includes RSD or reconstruction-induced contributions.
\end{itemize}

A particularly important structural feature of the ULPT formulation is that reconstruction affects only the displacement vector, leaving the combination $\delta_{\mathrm{Js}} + \delta_{\mathrm{b}}$ unchanged. As a result, the reconstructed density field remains identical to the pre-reconstruction field at linear order. This well-known property of linear theory is naturally and explicitly realized in our framework through the structure of Eq.~\eqref{eq:delta_rec_fourier_lag_J}.

Furthermore, as will be discussed in Sec.~\ref{sec:IR_resummed}, the displacement-mapping effect contributes directly to the nonlinear damping of the BAO signal. The effectiveness of reconstruction in restoring the BAO feature by mitigating such damping is also a well-established result. In our formulation, this behavior emerges in a natural and explicit manner from the mathematical structure of the equations, thereby providing a unified and transparent understanding of how reconstruction modifies the density field.

\section{Perturbation Theory Framework}
\label{sec:PTF}

In this section, we develop the perturbative structure of ULPT, focusing on the simplest case: dark matter in real space. The ULPT framework systematically decomposes the density fluctuation into two physically distinct components, the Jacobian deviation and the displacement-mapping effect, and evaluates their respective contributions order by order in perturbation theory. As a foundation, we begin with a brief review of the SPT kernels and the displacement field, followed by the construction and interpretation of perturbative solutions for the Jacobian deviation up to third order.

While our framework is general and applicable to RSDs, galaxy bias, and reconstruction within a unified treatment, we leave the detailed perturbative analysis of those effects for future work. The present section serves as a first step toward that goal, illustrating the underlying structure of the formulation in its most tractable setting.

For notational simplicity, we introduce the shorthand notation:
\begin{align}
    \int_{\boldsymbol{k}_{1\cdots n} = \boldsymbol{k}}
    &\equiv \int \frac{d^3k_1}{(2\pi)^3} \cdots \int \frac{d^3k_n}{(2\pi)^3}
    (2\pi)^3 \delta_{\rm D}\left(\boldsymbol{k} - \boldsymbol{k}_{1\cdots n} \right),
    \label{eq:k_sum_def} \\
    \boldsymbol{k}_{1\cdots n} &= \boldsymbol{k}_1 + \cdots + \boldsymbol{k}_n.
    \label{eq:k_sum}
\end{align}

\subsection{Standard Perturbative Treatments}

\subsubsection{Standard Perturbation Theory}

In the case of dark matter, the perturbative expansion of the density fluctuation is already well established at all orders within the framework of SPT. For a comprehensive review, see, for example, Ref.~\cite{Bernardeau:2001qr}.

In SPT, the density fluctuation is expanded as
\begin{equation}
    \delta(\boldsymbol{x}) = \sum_{n=1}^{\infty} \delta^{(n)}(\boldsymbol{x}),
    \label{eq:spt_expansion}
\end{equation}
where $\delta^{(n)}(\boldsymbol{x}) = \mathcal{O}([\delta^{(1)}]^n)$ represents the $n$-th order contribution in terms of the linear density field.

In linear theory, the dark matter density fluctuation can be separated into spatial and temporal components:
\begin{equation}
\delta^{(1)}(\boldsymbol{x}, z) = D(z)\, \delta^{(1)}_0(\boldsymbol{x}),
\end{equation}
where \( D(z) \) is the linear growth factor that governs the time evolution of the density fluctuation, and \( \delta^{(1)}_0(\boldsymbol{x}) \equiv \delta^{(1)}(\boldsymbol{x}, z=0) \) represents its value at the present time.

In Fourier space, the $n$-th order fluctuation is expressed as
\begin{equation}
    \widetilde{\delta}^{(n)}(\boldsymbol{k}) = \int_{\boldsymbol{k}_{1\cdots n} = \boldsymbol{k}}
    F_n(\boldsymbol{k}_1, \ldots, \boldsymbol{k}_n) \prod_{i=1}^n \widetilde{\delta}^{(1)}(\boldsymbol{k}_i),
    \label{eq:delta_n_fourier}
\end{equation}
where $F_n$ denotes the $n$-th order kernel function that encapsulates the nonlinear coupling of modes. In linear theory, $F_1 = 1$. For any order $n\geq2$, the kernel functions $F_{n}$ can be systematically computed using well-established recursion relations.

Similar to the case of the density fluctuation, the divergence of the velocity field in comoving coordinates, defined as \( \theta = \nabla \cdot \boldsymbol{v} \), can also be expanded perturbatively. The \( n \)-th order contribution in Fourier space is given by
\begin{equation}
    \widetilde{\theta}^{(n)}(\boldsymbol{k}) = -H f \int_{\boldsymbol{k}_{1\cdots n} = \boldsymbol{k}}
    G_n(\boldsymbol{k}_1, \ldots, \boldsymbol{k}_n) \prod_{i=1}^n \widetilde{\delta}^{(1)}(\boldsymbol{k}_i),
    \label{eq:theta_n_fourier}
\end{equation}
where \( G_1 = 1 \) and \( f = d\ln D / d\ln a \) is the linear growth rate function, with \( a \) being the scale factor.

As a concrete example, the second-order kernel $F_2$ takes the form
\begin{equation}
    F_2(\boldsymbol{k}_1, \boldsymbol{k}_2) = \frac{5}{7} + \frac{1}{2}
    (\hat{\boldsymbol{k}}_1 \cdot \hat{\boldsymbol{k}}_2)\left( \frac{k_1}{k_2} + \frac{k_2}{k_1} \right)
    + \frac{2}{7}(\hat{\boldsymbol{k}}_1 \cdot \hat{\boldsymbol{k}}_2)^2.
    \label{eq:F2}
\end{equation}
where $\hat{\boldsymbol{k}} = \boldsymbol{k} / |\boldsymbol{k}|$.

A characteristic property of the nonlinear kernels $F_{n\geq2}$ and $G_{n\geq2}$ is that they vanish when the total wavevector vanishes, i.e., when $\boldsymbol{k}_1 + \cdots + \boldsymbol{k}_n = 0$. This condition ensures that the density fluctuation satisfies two fundamental statistical constraints expected for a fluctuation field:
\begin{align}
    \int d^3x\, \delta(\boldsymbol{x}) &= \widetilde{\delta}(\boldsymbol{k} = 0) \nonumber \\
    & = 0,
    \quad \text{(vanishing volume average)},
    \label{eq:volume_average} \\
    \langle \delta(\boldsymbol{x}) \rangle &= 0,
    \quad \text{(vanishing ensemble average)}.
    \label{eq:ensemble_average}
\end{align}
The same conditions apply to the velocity divergence field \( \theta \) as well.

The vanishing of the linear-order average in Eq.~\eqref{eq:ensemble_average} is consistent with the statistical properties predicted by inflationary models of the early universe~\cite{Starobinsky:1980te,Sato:1980yn,Guth:1980zm,Linde:1981mu,Albrecht:1982wi}. In contrast, the nonlinear terms require a more subtle condition: the vanishing of both the volume and ensemble averages implies that the kernel functions must satisfy $F_{n\geq2} = 0$ when the wavevector sum vanishes. Thus, the structure of the nonlinear kernels plays an essential role in preserving the statistical definition of fluctuations at all orders.

The kernel functions $F_n$ should be symmetric under permutations of their wavevector arguments $(\boldsymbol{k}_1, \ldots, \boldsymbol{k}_n)$. Therefore, even if the recursion relations yield non-symmetric expressions, they must be symmetrized via
\begin{equation}
     F_n^{(\mathrm{sym})}(\boldsymbol{k}_1, \ldots, \boldsymbol{k}_n) = \frac{1}{n!} \sum_{\sigma \in {\cal S}_n} F_n(\boldsymbol{k}_{\sigma(1)}, \ldots, \boldsymbol{k}_{\sigma(n)}),
\end{equation}
where ${\cal S}_n$ denotes the set of all $n!$ permutations of $n$ elements. This symmetrization applies equally to all nonlinear kernel functions discussed in this paper. For notational simplicity, however, we do not indicate this operation explicitly in the expressions that follow.

\subsubsection{Perturbative Expansion of the Displacement Field}

According to Eq.~\eqref{eq:dJ}, the Jacobian deviation $\delta_{\rm J}$ can be directly computed from the displacement field. The $n$-th order perturbative solution of the displacement field in Fourier space is given by~\cite{Matsubara:2015ipa}
\begin{align}
    \widetilde{\boldsymbol{\Psi}}^{(n)}(\boldsymbol{k})
= \frac{i}{n!}
& \int_{\boldsymbol{k}_{1\cdots n}=\boldsymbol{k}}\, \boldsymbol{L}_n(\boldsymbol{k}_1, \ldots, \boldsymbol{k}_n)
\, \prod_{i=1}^{n}\widetilde{\delta}^{(1)}(\boldsymbol{k}_i).
\label{eq:psi_n}
\end{align}

The $n$-th order kernel vector $\boldsymbol{L}_n$ can be decomposed into longitudinal and transverse components as follows:
\begin{align}
\boldsymbol{L}_n(\boldsymbol{k}_1, \ldots, \boldsymbol{k}_n)
= \frac{1}{k_{1\cdots n}^2} \bigg[ \,
& \boldsymbol{k}_{1\cdots n} \,
S_n(\boldsymbol{k}_1, \ldots, \boldsymbol{k}_n)
\nonumber \\
& + \,
\boldsymbol{k}_{1\cdots n} \times
\boldsymbol{T}_n(\boldsymbol{k}_1, \ldots, \boldsymbol{k}_n)
\bigg].
\label{eq:Ln}
\end{align}
where $S_n$ and $\boldsymbol{T}_n$ are the scalar (longitudinal) and vector (transverse) components, respectively. In linear theory, $S_1=0$ and $\boldsymbol{T}_1=0$. All higher-order components ($n\geq2$) can be systematically computed using the recursion relations derived in Ref.~\cite{Matsubara:2015ipa}.

To compute $\delta_{\rm J}$ up to third order, we require the kernel functions up to $\boldsymbol{L}_3$. It is known that the transverse components vanish at second order, i.e., $\boldsymbol{T}_2 = 0$. Furthermore, the third-order contribution to $\delta_{\rm J}$ also does not involve $\boldsymbol{T}_3$. Therefore, in this work, we focus exclusively on the longitudinal components $S_n$ for $n = 1, 2, 3$.

To explicitly write down the solutions for $S_n$ up to third order, we follow the definitions introduced in Ref.~\cite{Matsubara:2015ipa}. We first define the following two auxiliary functions, which capture key geometric features of the interacting wavevectors.

The first function
\begin{align}
U(\boldsymbol{k}_1, \boldsymbol{k}_2)
&= |\hat{\boldsymbol{k}}_1\times \hat{\boldsymbol{k}}_2|^2 = 1 - (\hat{\boldsymbol{k}}_1\cdot\hat{\boldsymbol{k}}_2)^2
\label{eq:Udef}
\end{align}
represents the squared magnitude of the cross product between two unit wavevectors. It quantifies their degree of non-alignment: the value vanishes when the vectors are parallel and reaches its maximum when they are orthogonal. This function provides a geometric measure of the angular dependence relevant to second-order mode coupling.

The second function
\begin{align}
V(\boldsymbol{k}_1, \boldsymbol{k}_2, \boldsymbol{k}_3)
&= \left|\hat{\boldsymbol{k}}_1 \cdot(\hat{\boldsymbol{k}}_2 \times \hat{\boldsymbol{k}}_3)\right|^2
\nonumber \\
&= 1
- (\hat{\boldsymbol{k}}_1 \cdot \hat{\boldsymbol{k}}_2)^2
- (\hat{\boldsymbol{k}}_2 \cdot \hat{\boldsymbol{k}}_3)^2
- (\hat{\boldsymbol{k}}_3 \cdot \hat{\boldsymbol{k}}_1)^2
\nonumber \\
&\quad + 2 \,(\hat{\boldsymbol{k}}_1 \cdot \hat{\boldsymbol{k}}_2)
(\hat{\boldsymbol{k}}_2 \cdot \hat{\boldsymbol{k}}_3)
(\hat{\boldsymbol{k}}_3 \cdot \hat{\boldsymbol{k}}_1)
\label{eq:Vdef}
\end{align}
represents the squared scalar triple product of three unit wavevectors and characterizes their non-coplanarity. It vanishes when the three vectors lie in a common plane and becomes large when they span a maximally three-dimensional configuration.

Using these $U$ and $V$ functions, the longitudinal components $S_n$ for $n=2,3$ are given by:
\begin{align}
    S_2(\boldsymbol{k}_1,\boldsymbol{k}_2) &= \frac{3}{7} U(\boldsymbol{k}_1,\boldsymbol{k}_2), \nonumber \\
    S_3(\boldsymbol{k}_1,\boldsymbol{k}_2,\boldsymbol{k}_3) &= \frac{5}{3} U(\boldsymbol{k}_1,\boldsymbol{k}_{23}) S_2(\boldsymbol{k}_2,\boldsymbol{k}_{3}) 
    - \frac{1}{3} V(\boldsymbol{k}_1,\boldsymbol{k}_2,\boldsymbol{k}_3),
\label{eq:Sn}
\end{align}
where $\boldsymbol{k}_{23}=\boldsymbol{k}_{2}+\boldsymbol{k}_{3}$.

\subsection{Jacobian Deviation}

\subsubsection{General Properties}
\label{sec:J_Gene}

The $n$-th order perturbative solution of the Jacobian deviation $\delta_{\rm J}$ in Fourier space can be written as
\begin{equation}
    \widetilde{\delta}_{\rm J}^{(n)}(\boldsymbol{k}) = \int_{\boldsymbol{k}_{1\cdots n} = \boldsymbol{k}} J_n(\boldsymbol{k}_1, \ldots, \boldsymbol{k}_n) \prod_{i=1}^{n} \widetilde{\delta}^{(1)}(\boldsymbol{k}_i),
    \label{eq:deltaJ_n_def}
\end{equation}
where $J_n$ denotes the nonlinear kernel characterizing $\delta_{\rm J}^{(n)}$. These kernel functions are newly introduced in this work. At linear order, $\delta_{\rm J}^{(1)} = \delta^{(1)}$, and hence $J_1 = 1$.

As with the standard SPT kernels $F_n$, the Jacobian kernels $J_n$ satisfy the condition
\begin{equation}
    J_n(\boldsymbol{k}_1, \ldots, \boldsymbol{k}_n) = 0 \quad \text{when} \quad \boldsymbol{k}_1 + \cdots + \boldsymbol{k}_n = 0.
\end{equation}
This property follows from the identity
\begin{equation}
    \int d^3x \, \delta(\boldsymbol{x}) = \int d^3q \, \delta_{\rm J}(\boldsymbol{q}),
    \label{eq:delta_integrated}
\end{equation}
and the ensemble average relation
\begin{equation}
    \langle \delta(\boldsymbol{x}) \rangle = \langle \delta_{\rm J}(\boldsymbol{q}) \rangle.
    \label{eq:deltaJ_ensemble}
\end{equation}

Equation~\eqref{eq:delta_integrated} is straightforwardly derived from Eq.~\eqref{eq:delta_from_j}. In deriving Eq.~\eqref{eq:deltaJ_ensemble}, we use the translational invariance of ensemble averages. Specifically, statistical quantities defined in a statistical ensemble can, without loss of generality, be evaluated at the origin by exploiting spatial homogeneity:
\begin{equation}
    \langle \delta(\boldsymbol{x}) \rangle = \langle \delta(\boldsymbol{x} = \boldsymbol{0}) \rangle.
\end{equation}
Therefore, the ensemble average of $\delta(\boldsymbol{x})$ becomes
\begin{align}
    \langle \delta(\boldsymbol{x}) \rangle
    &= \left\langle\int d^3q \, \delta_{\rm J}(\boldsymbol{q} = \boldsymbol{0}) \, \delta_{\mathrm{D}}(\boldsymbol{x} - \boldsymbol{q} - \boldsymbol{\Psi}(\boldsymbol{q} = \boldsymbol{0})) \right\rangle\nonumber \\
    &= \langle \delta_{\rm J}(\boldsymbol{q} = \boldsymbol{0}) \rangle,
\end{align}
which confirms Eq.~\eqref{eq:deltaJ_ensemble}.

\subsubsection{Direct Calculations from the Jacobian}

The perturbative solutions for the Jacobian deviation $\delta_{\rm J}$ can be directly obtained by substituting the perturbative expressions for the displacement field $\boldsymbol{\Psi}$, given in Eqs.~\eqref{eq:psi_n}--\eqref{eq:Sn}, into the definition of the Jacobian deviation in Eq.~\eqref{eq:dJ}. The second- and third-order contributions are given by
\begin{align}
    \delta_{\rm J}^{(2)}(\boldsymbol{q}) &= -\Psi^{(2)}_{i,i}(\boldsymbol{q}) \nonumber \\
    &\quad - \frac{1}{2} \left[\Psi^{(1)}_{i,i}(\boldsymbol{q}) \Psi^{(1)}_{j,j}(\boldsymbol{q})
    - \Psi^{(1)}_{i,j}(\boldsymbol{q}) \Psi^{(1)}_{j,i}(\boldsymbol{q}) \right], 
    \label{eq:dJ2}
\end{align}
and
\begin{align}
    \delta_{\rm J}^{(3)}(\boldsymbol{q}) &= -\Psi^{(3)}_{i,i}(\boldsymbol{q}) \nonumber \\
    &\quad - \frac{1}{2} \big[ \Psi^{(2)}_{i,i}(\boldsymbol{q}) \Psi^{(1)}_{j,j}(\boldsymbol{q})
    + \Psi^{(1)}_{i,i}(\boldsymbol{q}) \Psi^{(2)}_{j,j}(\boldsymbol{q}) \nonumber \\
    &\qquad\quad - \Psi^{(2)}_{i,j}(\boldsymbol{q}) \Psi^{(1)}_{j,i}(\boldsymbol{q})
    - \Psi^{(1)}_{i,j}(\boldsymbol{q}) \Psi^{(2)}_{j,i}(\boldsymbol{q}) \big] \nonumber \\
    &\quad - \frac{1}{6} \epsilon_{ijk} \epsilon_{lmn}
    \Psi^{(1)}_{i,l}(\boldsymbol{q}) \Psi^{(1)}_{j,m}(\boldsymbol{q}) \Psi^{(1)}_{k,n}(\boldsymbol{q}).
    \label{eq:dJ3}
\end{align}
Importantly, the contribution from the third-order displacement field $\boldsymbol{\Psi}^{(3)}$ enters only through its divergence, $\nabla \cdot \boldsymbol{\Psi}^{(3)}$, indicating that only the longitudinal component of $\boldsymbol{\Psi}^{(3)}$ contributes to $\delta_{\rm J}^{(3)}$.

Using the nonlinear kernel vectors $\boldsymbol{L}_n$ up to third order, the corresponding second- and third-order kernel functions for the Jacobian deviation can be derived as
\begin{align}
    J_2(\boldsymbol{k}_1, \boldsymbol{k}_2) &= -\frac{2}{7} U(\boldsymbol{k}_1, \boldsymbol{k}_2), 
    \label{eq:J2} \\
    J_3(\boldsymbol{k}_1, \boldsymbol{k}_2, \boldsymbol{k}_3) &= -\frac{2}{9} U(\boldsymbol{k}_1, \boldsymbol{k}_{23}) S_2(\boldsymbol{k}_2, \boldsymbol{k}_3) + \frac{1}{9} V(\boldsymbol{k}_1, \boldsymbol{k}_2, \boldsymbol{k}_3), 
    \label{eq:J3}
\end{align}
where $U$ and $V$ are defined in Eqs.~\eqref{eq:Udef} and~\eqref{eq:Vdef}.

As expected, the kernel functions vanish when the total wavevector vanishes:
\begin{align}
    J_2(\boldsymbol{k}_1, \boldsymbol{k}_2) &= 0 \quad \text{for } \boldsymbol{k}_{12} = \boldsymbol{0}, \\
    J_3(\boldsymbol{k}_1, \boldsymbol{k}_2, \boldsymbol{k}_3) &= 0 \quad \text{for } \boldsymbol{k}_{123} = \boldsymbol{0}. \label{eq:J_vanish}
\end{align}
This cancellation in $J_3$ arises from the fact that both terms in Eq.~\eqref{eq:J3} individually vanish when $\boldsymbol{k}_{123} = 0$:
\begin{align}
    U(\boldsymbol{k}_1, \boldsymbol{k}_{23})S_2(\boldsymbol{k}_2, \boldsymbol{k}_{3}) 
    &= 0 \quad \text{for } \boldsymbol{k}_{123} = \boldsymbol{0}, \\
    V(\boldsymbol{k}_1, \boldsymbol{k}_2, \boldsymbol{k}_3) &= 0 \quad \text{for } \boldsymbol{k}_{123} = \boldsymbol{0}. \label{eq:UV_vanish}
\end{align}

Although the full third-order density fluctuation $\delta^{(3)}(\boldsymbol{x})$ typically has a complicated structure, the contribution from the Jacobian deviation alone, excluding the displacement-mapping effect, turns out to be remarkably simple. It is fully characterized by only two terms: $U(\boldsymbol{k}_1, \boldsymbol{k}_2) = |\hat{\boldsymbol{k}}_1 \times \hat{\boldsymbol{k}}_2|^2$ and $V(\boldsymbol{k}_1, \boldsymbol{k}_2, \boldsymbol{k}_3) = |\hat{\boldsymbol{k}}_1 \cdot (\hat{\boldsymbol{k}}_2 \times \hat{\boldsymbol{k}}_3)|^2$, involving only inner and cross products of the wavevectors. This structural simplicity is a noteworthy feature of our formulation.

\subsubsection{Alternative Construction from the Relation to $\delta$}

In our formulation, the dark matter density contrast $\delta$ is constructed from the Jacobian deviation $\delta_{\rm J}$ and the displacement vector $\boldsymbol{\Psi}$. At each perturbative order, the left-hand and right-hand sides of this relationship remain equal. This identity offers an alternative method to derive the $n$th-order $\delta_{\rm J}$.

From Eq.~\eqref{eq:d_dJ_Psi}, the second- and third-order density contrasts in Eulerian space are given by
\begin{align}
    \delta^{(2)}(\boldsymbol{x}) &= \delta_{\rm J}^{(2)}(\boldsymbol{x})
    - \partial_i \left[\Psi_i^{(1)}(\boldsymbol{x}) \, \delta_{\rm J}^{(1)}(\boldsymbol{x}) \right],
    \label{eq:delta2_alt}
\end{align}
and
\begin{align}
    \delta^{(3)}(\boldsymbol{x}) &= \delta_{\rm J}^{(3)}(\boldsymbol{x})
    - \partial_i\left[ \Psi_i^{(2)}(\boldsymbol{x}) \, \delta_{\rm J}^{(1)}(\boldsymbol{x}) \right]
    \notag \\
    &\quad - \partial_i\left[\Psi_i^{(1)}(\boldsymbol{x}) \, 
    \delta_{\rm J}^{(2)}(\boldsymbol{x}) \right]
    \notag \\
    &\quad + \frac{1}{2}\partial_i\partial_j 
    \left[\Psi_i^{(1)}(\boldsymbol{x}) \Psi_j^{(1)}(\boldsymbol{x}) \, \delta_{\rm J}^{(1)}(\boldsymbol{x}) \right].
    \label{eq:delta3_alt}
\end{align}

These relations allow us to express the nonlinear kernel $J_n$ of the Jacobian deviation in terms of the standard density kernels $F_n$ and the displacement kernels $\boldsymbol{L}_n$ as
\begin{align}
    J_2(\boldsymbol{k}_1, \boldsymbol{k}_2) &= F_2(\boldsymbol{k}_1, \boldsymbol{k}_2)
    - \frac{1}{2}
    \boldsymbol{k}_{12} \cdot \left[  \boldsymbol{L}_1(\boldsymbol{k}_1)+ \boldsymbol{L}_1(\boldsymbol{k}_2)\right],
    \label{eq:J2_from_F2_L1}
\end{align}
and
\begin{align}
    & J_3(\boldsymbol{k}_1, \boldsymbol{k}_2, \boldsymbol{k}_3) \nonumber \\
    &= F_3(\boldsymbol{k}_1, \boldsymbol{k}_2, \boldsymbol{k}_3)
    - \frac{1}{2}\, \left[\boldsymbol{k}_{123} \cdot \boldsymbol{L}_2(\boldsymbol{k}_1, \boldsymbol{k}_2)  \right]
    \nonumber \\
    & - \left[\boldsymbol{k}_{123} \cdot \boldsymbol{L}_1(\boldsymbol{k}_3)  \right] 
    \, F_2(\boldsymbol{k}_1, \boldsymbol{k}_2) \nonumber \\
    & + \frac{1}{2}\left[\boldsymbol{k}_{123} \cdot \boldsymbol{L}_1(\boldsymbol{k}_3)  \right] 
    \, \left[ \boldsymbol{k}_{12}\cdot \left( \boldsymbol{L}_1(\boldsymbol{k}_1)  
    +  \boldsymbol{L}_1(\boldsymbol{k}_2)\right)\right]\nonumber \\
    & - \frac{1}{2}\, \left[ \boldsymbol{k}_{123} \cdot \boldsymbol{L}_1(\boldsymbol{k}_1) \right]
                 \left[ \boldsymbol{k}_{123} \cdot \boldsymbol{L}_1(\boldsymbol{k}_2) \right].
    \label{eq:J3_from_F3_L}
\end{align}

As shown above, the nonlinear kernel $J_n$ can be constructed by directly substituting the expressions for $F_n$ and $L_n$, which are obtained from their respective recursion relations.

\subsection{Interpretation of Perturbative Contributions}

In this subsection, we provide a physical interpretation of the perturbative contributions discussed in the previous sections, with emphasis on their roles within the ULPT framework. For notational simplicity, we omit explicit Eulerian coordinate dependence; for example, we write $\delta^{(1)}$ instead of $\delta^{(1)}(\boldsymbol{x})$.

\subsubsection{Conventional Decomposition of Second- and Third-Order Terms}

It is well known that the second-order dark matter density contrast can be decomposed into three physically distinct components~\cite{Bouchet:1992xz,Baldauf:2012hs,Sherwin:2012,Schmittfull:2014tca}:
\begin{equation}
\delta^{(2)} = \frac{17}{21} \underbrace{[\delta^{(1)}]^2}_{\text{Nonlinear growth}}
- \underbrace{\Psi_i^{(1)} \partial_i \delta^{(1)}}_{\text{Shift term}}
+ \frac{2}{7} \underbrace{K^{(1)}_{ij} K^{(1)}_{ij}}_{\text{Tidal term}},
\label{eq:delta2_terms}
\end{equation}
where the first term represents the nonlinear growth of matter density fluctuations, corresponding to the spherical collapse contribution in the absence of tidal effects. The second term, known as the \emph{shift term}, accounts for the convective transport of matter induced by large-scale displacements. The third term is the \emph{tidal term}, involving the tidal tensor defined by
\begin{equation}
K^{(1)}_{ij} \equiv D_{ij} \delta^{(1)},
\label{eq:tidal_tensor}
\end{equation}
with the differential operator
\begin{equation}
D_{ij} \equiv \left( \frac{\partial_i \partial_j}{\partial^2} - \frac{1}{3} \delta_{ij} \right),
\label{eq:tidal_operator}
\end{equation}
where $\delta_{ij}$ is the Kronecker delta and $\partial^2 = \delta_{ij} \partial_i \partial_j$ is the Laplacian operator.

At third order, the density contrast, excluding the shift-type contributions, can be characterized by four distinct scalar operators~\cite{Mirbabayi:2014zca,Desjacques:2016bnm}:
\begin{equation}
[\delta^{(1)}]^3, \quad \delta^{(1)} K_{ij}^{(1)} K_{ij}^{(1)}, \quad K_{ij}^{(1)} K_{jk}^{(1)} K_{ki}^{(1)}, \quad O_{\mathrm{td}}^{(3)},
\label{eq:third_order_operators}
\end{equation}
where $O_{\mathrm{td}}^{(3)}$ is defined by
\begin{equation}
    O_{\mathrm{td}}^{(3)} = \frac{8}{21} K_{ij}^{(1)} D_{ij} \left( [\delta^{(1)}]^2 - \frac{3}{2} K_{kl}^{(1)} K_{kl}^{(1)} \right).
\label{eq:O_td_def}
\end{equation}

An alternative decomposition of the second- and third-order dark matter density contrast, equivalent to Eqs.~(\ref{eq:delta2_terms}) and~(\ref{eq:third_order_operators}), can be formulated using Galileon operators~\cite{Chan:2012jj,Assassi:2014fva}. In this approach, one defines the rescaled gravitational and velocity potentials as
\begin{equation}
\Phi_g \equiv \partial^{-2} \delta, \quad \Phi_v \equiv - (Hf)^{-1} \partial^{-2} \theta,
\end{equation}
where $\partial^{-2}$ denotes the inverse Laplacian. Using these potentials, the Galileon operators up to third order are constructed as follows:
\begin{align}
\mathcal{G}_2(\Phi_g) &\equiv
(\partial_{ij} \Phi_g)^2
- (\partial^2 \Phi_g)^2,
\label{eq:G2_def} \\
\mathcal{G}_3(\Phi_g) &\equiv
(\partial^2 \Phi_g)^3
+ 2\, \partial_{ij} \Phi_g \, \partial_{jk} \Phi_g \, \partial_{ki} \Phi_g
\nonumber \\
&\quad
- 3\, (\partial_{ij} \Phi_g)^2 \, \partial^2 \Phi_g,
\label{eq:G3_def} \\
\Gamma_3 &\equiv
\mathcal{G}_2(\Phi_g) - \mathcal{G}_2(\Phi_v).
\label{eq:Gamma3_def}
\end{align}
where we define $\partial_{ij} \equiv \partial_i \partial_j$. Note that in some conventions (e.g., Eq.~(C.18) of Ref.~\cite{Desjacques:2016bnm}), a prefactor of $-1/2$ is included in the definition of $\mathcal{G}_3$.

At second order, the Galileon operator can be expressed in terms of the linear density and tidal tensor as
\begin{equation}
\mathcal{G}_2^{(2)} = -\frac{2}{3}[\delta^{(1)}]^2 + K_{ij}^{(1)} K_{ij}^{(1)}.
\label{eq:G_K_2}
\end{equation}
Substituting this into the standard second-order expression in Eq.~\eqref{eq:delta2_terms}, the dark matter density contrast can be rewritten as
\begin{equation}
\delta^{(2)} = \underbrace{[\delta^{(1)}]^2}_{\text{Nonlinear growth}}
- \underbrace{\Psi^{(1)}_i \, \partial_i \delta^{(1)}}_{\text{Shift}}
+ \frac{2}{7} \underbrace{\mathcal{G}^{(2)}_2}_{\text{Galileon}}.
\label{eq:delta2_terms_G}
\end{equation}
This form highlights the geometric structure of second-order contributions and provides a compact representation in terms of scalar invariants derived from the gravitational potential.

At third order, after removing the shift contribution, the density fluctuation can be fully characterized by the following four scalar operators:
\begin{equation}
[\delta^{(1)}]^3, \quad
\delta^{(1)} \mathcal{G}_2^{(2)}, \quad
\mathcal{G}_3^{(3)}, \quad
\Gamma_3^{(3)},
\label{eq:third_order_operators_G}
\end{equation}
These operators form a complete basis for describing the non-convective part of the third-order dark matter fluctuation. Their correspondence with the individual terms in Eq.~(\ref{eq:third_order_operators}) is given by the following identities:
\begin{align}
\delta^{(1)} \mathcal{G}_2^{(2)} 
&= -\frac{2}{3} [\delta^{(1)}]^3 + \delta^{(1)} K_{ij}^{(1)} K_{ij}^{(1)}, \nonumber \\
\mathcal{G}_3^{(3)} 
&= 2 K_{ij}^{(1)} K_{jk}^{(1)} K_{ki}^{(1)} 
- \delta^{(1)} K_{ij}^{(1)} K_{ij}^{(1)} + \frac{2}{9} [\delta^{(1)}]^3, \nonumber \\
\Gamma_3^{(3)} 
&= \frac{8}{21} \left[ \frac{21}{8}Q_{\mathrm{td}}^{(3)} - \frac{2}{3} [\delta^{(1)}]^3 + \delta^{(1)} K_{ij}^{(1)} K_{ij}^{(1)} \right].
\label{eq:G_K_3}
\end{align}

These considerations form the basis of the conventional approach to modeling galaxy bias, in which the density contrast is expanded in terms of local and nonlocal Lagrangian operators constructed from $\delta^{(1)}$ and the tidal tensor $K^{(1)}_{ij}$, while excluding shift-type contributions. Each operator is then multiplied by a corresponding bias parameter that encapsulates the tracer response to the underlying matter distribution.

Notably, these standard bias operators can be systematically mapped onto Galileon invariants such as $\mathcal{G}_2$ and $\mathcal{G}_3$, which provide an alternative and geometrically motivated basis for describing nonlinear gravitational evolution. This connection implies that the bias expansion may be equivalently formulated in terms of scalar combinations of the gravitational potential and its derivatives.

For a comprehensive review of Lagrangian bias modeling and its operator basis, see Ref.~\cite{Desjacques:2016bnm}.

\subsubsection{Galileon-Based Decomposition in the ULPT Framework}
\label{sec:interp_J}

In this subsection, we interpret the conventional decomposition of second- and third-order density fluctuations in terms of Galileon-type operators, within the ULPT framework developed in this work.

In the ULPT framework, the second-order density contrast is decomposed into two physically distinct contributions: the Jacobian deviation and the displacement-mapping effect. Rewriting Eq.~\eqref{eq:delta2_alt}, we have
\begin{equation}
    \delta^{(2)} = \underbrace{\delta^{(2)}_{\rm J}}_{\text{Jacobian deviation}} \, + \, \underbrace{- \partial_i \left(  \Psi^{(1)}_i\, \delta_{\rm J}^{(1)}\right)}_{\text{Displacement-mapping}}. \label{eq:delta2_decomp}
\end{equation}

Each term can be written explicitly as
\begin{align}
    \delta^{(2)}_{\rm J} &= \frac{2}{7}\, \mathcal{G}_2^{(2)}, \label{eq:delta2_J_explicit} \\
    -\partial_i \left(\Psi^{(1)}_i\, \delta_{\rm J}^{(1)}  \right) &= [\delta^{(1)}]^2 - \Psi^{(1)}_i \partial_i \delta^{(1)}, \label{eq:delta2_mapping}
\end{align}
where we have used the first-order identities $\delta^{(1)} = -\partial_i \Psi^{(1)}_i$ and $\delta^{(1)}=\delta^{(1)}_{\rm J}$.

It is noteworthy that at second order, the Jacobian deviation depends solely on the Galileon operator $\mathcal{G}_2^{(2)}$. This stems from the fact that the Fourier-space kernel $J_2$ depends exclusively on $U = |\hat{\boldsymbol{k}}_1 \times \hat{\boldsymbol{k}}_2|^2$, which corresponds to the scalar invariant appearing in $\mathcal{G}_2$. Consequently, the quadratic density term $[\delta^{(1)}]^2$ does not appear as an independent component in $\delta_{\rm J}^{(2)}$; rather, it always enters in the specific combination $-\frac{2}{3}[\delta^{(1)}]^2 + K^{(1)}_{ij} K^{(1)}_{ij}$ that defines the Galileon operator.

The displacement-mapping term in Eq.~\eqref{eq:delta2_mapping} fully contains the shift contribution and also gives rise to a component that resembles nonlinear growth. However, this latter component arises solely through the identity $\partial_i \Psi^{(1)}_i = -\delta^{(1)}$ and should therefore not be regarded as a genuine nonlinear growth effect. Instead, it represents an apparent contribution that is not independent of the shift term, but rather an intrinsic part of the displacement-mapping mechanism. We therefore interpret the convective transport of density fluctuations induced by large-scale displacements not as a standalone shift term, but as a phenomenon that should be described entirely by the displacement-mapping mechanism.

At third order, the Jacobian deviation can again be decomposed into two distinct contributions, corresponding to the $U\, S_2$ and $V$ parts of the kernel $J_3$ in Eq.~\eqref{eq:J3}. We denote these contributions by $\delta^{(3)}_{ {\rm J},U}$ and $\delta^{(3)}_{ {\rm J},V}$, respectively. Their explicit forms are given by
\begin{align}
    \delta^{(3)}_{{\rm J}, U}
    & = - \frac{2}{9} \int \frac{d^3k_1}{(2\pi)^3} \int \frac{d^3k_2}{(2\pi)^3} \int \frac{d^3k_3}{(2\pi)^3}
    e^{i(\boldsymbol{k}_1 + \boldsymbol{k}_2 + \boldsymbol{k}_3)\cdot \boldsymbol{x}} \nonumber \\
    & \quad \times  U(\boldsymbol{k}_1, \boldsymbol{k}_{23})\, S_2(\boldsymbol{k}_2, \boldsymbol{k}_3) 
    \tilde{\delta}^{(1)}(\boldsymbol{k}_1)
    \tilde{\delta}^{(1)}(\boldsymbol{k}_2)
    \tilde{\delta}^{(1)}(\boldsymbol{k}_3),
\end{align}
and
\begin{align}
    \delta^{(3)}_{{\rm J}, V}
    & = \frac{1}{9} \int \frac{d^3k_1}{(2\pi)^3} \int \frac{d^3k_2}{(2\pi)^3} \int \frac{d^3k_3}{(2\pi)^3}
    e^{i(\boldsymbol{k}_1 + \boldsymbol{k}_2 + \boldsymbol{k}_3)\cdot \boldsymbol{x}} \nonumber \\
    & \quad \times V(\boldsymbol{k}_1, \boldsymbol{k}_2, \boldsymbol{k}_3)
    \tilde{\delta}^{(1)}(\boldsymbol{k}_1)
    \tilde{\delta}^{(1)}(\boldsymbol{k}_2)
    \tilde{\delta}^{(1)}(\boldsymbol{k}_3),
\end{align}
where $U$ and $V$ are the scalar functions defined in Eqs.~\eqref{eq:Udef} and \eqref{eq:Vdef}, respectively.

These functions $U$ and $V$ are directly related to the Galileon-type operators. Specifically, the $U$-type contribution is given by
\begin{align}
    \delta^{(3)}_{ {\rm J},U} = \frac{1}{6}\, \Gamma^{(3)}_3, \label{eq:delta3_JU}
\end{align}
while the $V$-type contribution takes the form
\begin{align}
    \delta^{(3)}_{ {\rm J},V} = \frac{1}{9}\, \mathcal{G}^{(3)}_3. \label{eq:delta3_JV}
\end{align}
Thus, in our formulation, the $U$-type and $V$-type contributions to the third-order density fluctuation naturally map onto the third-order Galileon operators $\Gamma^{(3)}_3$ and $\mathcal{G}^{(3)}_3$.

Accordingly, the full third-order Jacobian deviation is expressed as
\begin{equation}
    \delta_{\rm J}^{(3)} =  \frac{1}{6} \Gamma^{(3)}_3 +  \frac{1}{9} \mathcal{G}_{3}^{(3)}.
    \label{eq:delta3_J_total}
\end{equation}

The third-order displacement-mapping contribution can be computed as
\begin{align}
    & -\partial_i (\Psi^{(1)}_i \delta_{\rm J}^{(2)})
    -\partial_i (\Psi^{(2)}_i \delta_{\rm J}^{(1)})
    + \frac{1}{2}\partial_i\partial_j \left( \Psi^{(1)}_i\Psi^{(1)}_j\delta_{\rm J}^{(1)} \right)
    \nonumber \\
    & = \frac{4}{7} \delta^{(1)} \mathcal{G}_2^{(2)} + [\delta^{(1)}]^3
    + \mbox{[shift-type terms]},
    \label{eq:delta3_mapping}
\end{align}
where the terms involving spatial derivatives of the Jacobian deviation, such as $\partial_i \delta_{\rm J}^{(1)}$ and $\partial_i \delta_{\rm J}^{(2)}$, are collectively referred to as shift-type terms.

By adding Eqs.~\eqref{eq:delta3_J_total} and \eqref{eq:delta3_mapping}, the full third-order dark matter density fluctuation becomes
\begin{align}
\delta^{(3)} =\;&
\frac{1}{9} \mathcal{G}^{(3)}_3
+ \frac{1}{6} \Gamma^{(3)}_3 
 + \frac{4}{7} \delta^{(1)} \mathcal{G}_2^{(2)}
+ [\delta^{(1)}]^3 \nonumber \\
& + \text{[shift-type terms]}.
\end{align}
Using the relations given in Eqs.~\eqref{eq:G_K_2} and~\eqref{eq:G_K_3}, this expression can also be written as
\begin{align}
    \delta^{(3)} & = \frac{341}{567} [\delta^{(1)}]^3 + \frac{11}{21} \delta^{(1)} K^{(1)}_{ij} K^{(1)}_{ij} 
    \nonumber \\
   & \quad + \frac{2}{9} K^{(1)}_{ij} K^{(1)}_{jk} K^{(1)}_{ki} + \frac{1}{6} \mathcal{O}_{\rm td}^{(3)}
    \nonumber \\
& + \text{[shift-type terms]}.
\end{align}
which coincides with Eq.~(B.46) of Ref.~\cite{Desjacques:2016bnm}.

In our formulation, the intrinsic third-order density fluctuations arise solely from the two Galileon-type operators, $\mathcal{G}_3^{(3)}$ and $\Gamma_3^{(3)}$. In contrast, the remaining two terms, $[\delta^{(1)}]^3$ and $\delta^{(1)}\mathcal{G}_2^{(2)}$, are interpreted as part of the displacement-mapping effect.

Galileon operators possess the property that their ensemble averages vanish~\cite{Chan:2012jj,Assassi:2014fva}. Specifically, we have
\begin{equation}
\langle \mathcal{G}_2^{(2)} \rangle = \langle \mathcal{G}_3^{(3)} \rangle = \langle \Gamma_3^{(3)} \rangle = 0.
\end{equation}
As discussed in Sec.~\ref{sec:J_Gene}, the Jacobian deviation satisfies $\langle \delta_{\rm J} \rangle = 0$ at any order in perturbation theory. It is therefore natural that the Jacobian deviation is fully described by linear combinations of Galileon operators.

On the other hand, the contributions that arise from the displacement-mapping mechanism, such as $[\delta^{(1)}]^2$, $[\delta^{(1)}]^3$, and $\delta^{(1)}\mathcal{G}_2^{(2)}$, do not vanish under ensemble averaging when considered individually.

Thus, the ULPT framework naturally separates contributions with vanishing ensemble averages from those that do not, providing a structurally transparent classification of physical effects in the density field.

\section{Power Spectrum}
\label{sec:power-spectrum}

In this section, we develop theoretical predictions for the power spectrum based on the ULPT framework proposed in this work. We first present the general expression for the power spectrum, and then examine how IR effects, including large-scale bulk flows, are treated within our formulation. The total contribution to the power spectrum is systematically decomposed into physically distinct components, corresponding to the Jacobian deviation and the displacement-mapping effect introduced in previous sections. This analysis serves as a concrete application of our formalism and lays the groundwork for further extensions to include galaxy bias, redshift-space distortions, and reconstruction effects.

\subsection{Unified Formulation of the Power Spectrum}

\subsubsection{Derivation from the ULPT Framework}

The power spectrum of the reconstructed galaxy density field is defined by
\begin{equation}
    \langle \widetilde{\delta}_{\rm gs,rec}(\boldsymbol{k})\,\widetilde{\delta}_{\rm gs,rec}(\boldsymbol{k}') \rangle
    = (2\pi)^3 \delta_{\rm D}(\boldsymbol{k} + \boldsymbol{k}')\, P_{\rm gs,rec}(\boldsymbol{k}). \label{eq:def_power}
\end{equation}
Using Eq.~\eqref{eq:delta_rec_fourier_lag_J}, the power spectrum can be written as
\begin{align}
P_{\rm gs,rec}(\boldsymbol{k}) 
&= \int d^3 r_q\, e^{-i\boldsymbol{k}\cdot\boldsymbol{r}_q}
\Bigg\langle e^{-i \boldsymbol{k} \cdot [\boldsymbol{\Psi}_{\rm s,rec}(\boldsymbol{q}) 
- \boldsymbol{\Psi}_{\rm s,rec}(\boldsymbol{q}')]} \nonumber \\
&\quad \times 
 \left[ \delta_{\rm Js}(\boldsymbol{q}) + \delta_{\rm b}(\boldsymbol{q}) \right] 
\left[ \delta_{\rm Js}(\boldsymbol{q}') + \delta_{\rm b}(\boldsymbol{q}') \right] \Bigg\rangle. 
\label{eq:P_def}
\end{align}
Here, $\boldsymbol{r}_q \equiv \boldsymbol{q} - \boldsymbol{q}'$ denotes the separation vector between two Lagrangian positions.

Let $X$, $Y$, and $Y'$ be arbitrary statistical variables. The following identity for the ensemble average holds:
\begin{equation}
    \langle e^X Y Y' \rangle
    = \langle e^X \rangle 
    \left[  \langle e^X Y Y' \rangle_{\rm c} + \langle e^X Y \rangle_{\rm c} \langle e^X Y' \rangle_{\rm c}\right]
\label{eq:XYZ}
\end{equation}
where $\langle e^{X}\rangle$ is the moment-generating function of $X$, given by
\begin{equation}
    \langle e^X \rangle = \exp\left( \sum_{m=1}^{\infty} \frac{1}{m!} \langle X^m \rangle_{\rm c} \right), \label{eq:expX}
\end{equation}
and the subscript ``c'' denotes the connected part.

Applying this identity to the case
\begin{align}
    X &= -i \boldsymbol{k} \cdot [\boldsymbol{\Psi}_{\rm s,rec}(\boldsymbol{q}) 
    - \boldsymbol{\Psi}_{\rm s,rec}(\boldsymbol{q}')], \nonumber \\
    Y &= \delta_{\rm Js}(\boldsymbol{q}) + \delta_{\rm b}(\boldsymbol{q}), \nonumber \\
    Y' &= \delta_{\rm Js}(\boldsymbol{q}') + \delta_{\rm b}(\boldsymbol{q}'), 
    \label{eq:XYZ_def}
\end{align}
the moment-generating function becomes~\cite{Matsubara:2007wj}
\begin{equation}
\langle e^X \rangle = \exp\left[-\overline{\Sigma}_{\rm s,rec}(\boldsymbol{k}) 
+ \Sigma_{\rm s,rec}(\boldsymbol{k}, \boldsymbol{r}_q) \right], 
\label{eq:Sigma_def}
\end{equation}
where the exponent is defined as
\begin{align}
    &-\overline{\Sigma}_{\rm s,rec}(\boldsymbol{k}) + \Sigma_{\rm s,rec}(\boldsymbol{k}, \boldsymbol{r}_q) \nonumber\\
&= \sum_{m=2}^{\infty} \frac{1}{m!}\,
    \left\langle\,
      \bigl[-i \boldsymbol{k}\!\cdot\!
      \bigl(\boldsymbol{\Psi}_{\rm s,rec}(\boldsymbol{q})
           - \boldsymbol{\Psi}_{\rm s,rec}(\boldsymbol{q}')\bigr)\bigr]^{m}
    \right\rangle_{\!c}.
\label{eq:Sigma_expand}
\end{align}
Note that the expansion starts from $m=2$ due to $\langle \boldsymbol{\Psi}_{\rm s,rec} \rangle = 0$.

When considering the difference of a single displacement field, the exponent in Eq.~(\ref{eq:Sigma_expand}) vanishes at $\boldsymbol{r}_q = \boldsymbol{q}-\boldsymbol{q}'=0$. In this case, the variance $\overline{\Sigma}_{s,{\rm rec}}(\boldsymbol{k})$ is defined as the zero-separation limit of $\Sigma_{s,{\rm rec}}(\boldsymbol{k},\boldsymbol{r}_q)$:
\begin{equation}
    \overline{\Sigma}_{s,{\rm rec}}(\boldsymbol{k}) \equiv \Sigma_{s,{\rm rec}}(\boldsymbol{k}, \boldsymbol{r}_q = 0).
\end{equation}
This simplification, however, no longer holds when computing cross-power spectra between fields characterized by distinct displacement vectors, in which case the cancellation at $\boldsymbol{r}_q = 0$ is not guaranteed. See Sec.~\ref{sec:cross_P} for a detailed discussion.

We then define the remaining part of the integrand as
\begin{equation}
    \xi_{\rm Jgs,rec}(\boldsymbol{r})
    \equiv \langle e^X Y Y' \rangle_c + \langle e^X Y \rangle_c \langle e^X Y' \rangle_c.
\label{eq:xi_J_def}
\end{equation}
Using the relations $\langle X \rangle = \langle Y \rangle = \langle Y'\rangle=0$, we obtain
\begin{align}
    \xi_{\rm Jgs,rec}(\boldsymbol{r}) & =
    \langle Y Y' \rangle_c +
    \langle (e^X-1) Y Y' \rangle_c \nonumber \\
    & \quad +
    \langle (e^X-1) Y \rangle_c \, \langle (e^X-1) Y' \rangle_c .
\end{align}
which shows that $\xi_{\rm Jgs,rec}$ consists of both the self-correlations of the Jacobian deviation with bias
$Y = \delta_{\rm Js}+\delta_{\rm b}$ and its cross-correlations with the displacement-mapping effect.

Combining Eq.~\eqref{eq:Sigma_def} with Eq.~\eqref{eq:xi_J_def}, we obtain the general ULPT expression for the power spectrum:
\begin{equation}
    P_{\rm gs,rec}(\boldsymbol{k}) = e^{-\overline{\Sigma}_{\rm s,rec}(\boldsymbol{k})}
    \int d^3 r_q \, e^{-i \boldsymbol{k} \cdot \boldsymbol{r}_q} \,
    e^{\Sigma_{\rm s,rec}(\boldsymbol{k}, \boldsymbol{r}_q)} \,
    \xi_{\rm Jgs,rec}(\boldsymbol{r}_q).
\label{eq:P_general}
\end{equation}
This constitutes one of the main results of this work. It demonstrates that the power spectrum, whether for dark matter, biased tracers, redshift-space distortions, or reconstruction, can always be computed within the ULPT framework in this unified form.

The two fundamental building blocks of the ULPT power spectrum are defined as follows:

\begin{itemize}

    \item \textbf{Displacement-mapping factor}
    $e^{ - \overline{\Sigma}_{\rm s,rec}(\boldsymbol{k}) + \Sigma_{\rm s,rec}(\boldsymbol{k},\boldsymbol{r})}$:
    the exponential prefactor in Eq.~\eqref{eq:P_general}, which originates solely from the displacement-mapping effect.
    It encapsulates the impact of large-scale bulk flows and long-wavelength modes by modulating the clustering pattern through convective coordinate remapping.
    Importantly, it is statistically uncorrelated with the intrinsic source fields and therefore isolates the IR-sensitive contribution in a non-perturbative manner.

    \item \textbf{Source correlation function} $\xi_{\rm Jgs,rec}(\boldsymbol{r}_q)$: the two-point statistics of the composite intrinsic field $Y = [\delta_{\rm Js} + \delta_{\rm b}]$, evaluated with the displacement weighting $e^{X}$. This definition includes not only the self-correlations of $Y$ but also its cross-correlations with the displacement field through $X$. At linear order, $\xi_{\rm Jgs,rec}$ receives contributions solely from the Jacobian deviation and the biased fluctuation, i.e., from the two-point correlations of the first-order pieces of $Y$. At higher orders, additional contributions arise from (i) nonlinear self-correlations of the intrinsic fields and (ii) cross-correlations with the displacement field generated by the perturbative expansion of $e^{X}$. Within the ULPT framework, $\xi_{\rm Jgs,rec}$ serves as the intrinsic ``source'' component of the power-spectrum expression and represents the only place where galaxy-bias parameters appear.

\end{itemize}

\subsubsection{Fourier-Space Representation}

Expanding the exponential as
\begin{equation}
e^{\Sigma_{\rm s,rec}(\boldsymbol{k}, \boldsymbol{r}_q)} = \sum_{n=0}^{\infty} \frac{1}{n!} [\Sigma_{\rm s,rec}(\boldsymbol{k}, \boldsymbol{r}_q)]^n, \label{eq:exp_Sigma_expand}
\end{equation}
we obtain the Fourier-space expression:
\begin{align}
P_{\rm gs,rec}(\boldsymbol{k})
&= e^{-\overline{\Sigma}_{\rm s,rec}(\boldsymbol{k})}\sum_{n=0}^{\infty} \frac{1}{n!} \nonumber \\
& \times \int_{\boldsymbol{k}_{1}+\cdots+\boldsymbol{k}_{n+1}= \boldsymbol{k}}
\left[  \prod_{i=1}^{n} \widetilde{\Sigma}_{\rm s,rec}(\boldsymbol{k}, \boldsymbol{k}_i)\right]\,
P_{\rm Jgs,rec}(\boldsymbol{k}_{n+1}).
\label{eq:P_expanded}
\end{align}
where
\begin{align}
    \widetilde{\Sigma}_{\rm s,rec}(\boldsymbol{k}, \boldsymbol{k}_i) &= \int d^3 r_q\, e^{-i \boldsymbol{k}_i \cdot \boldsymbol{r}_q} \Sigma_{\rm s,rec}(\boldsymbol{k}, \boldsymbol{r}_q), 
    \label{eq:Sigma_k_ki} \\
P_{\rm Jgs,rec}(\boldsymbol{k}) &= \int d^3 r_q\, e^{-i \boldsymbol{k} \cdot \boldsymbol{r}_q} \xi_{\rm Jgs,rec}(\boldsymbol{r}_q). \label{eq:PJ_def}
\end{align}
We refer to \( P_{\rm Jgs,rec} \) as the \emph{source power spectrum}, as it is defined as the Fourier transform of the source correlation function \( \xi_{\rm Jgs,rec} \).

As seen in Eq.~\eqref{eq:P_expanded}, this expression involves an infinite series of mode-coupling integrals in Fourier space, rendering direct computation practically intractable. A key advantage of ULPT is that it allows for efficient evaluation of the power spectrum by performing the convolution integrals in configuration space, thereby resumming the infinite series generated by the displacement-mapping factor.

\subsection{Perturbative Expansion}

Our guiding principle for performing perturbative calculations within the ULPT framework is as follows: at each order in perturbation theory, we first reproduce the standard solution from Eulerian perturbation theory. Any additional contributions beyond this standard result arise solely from the displacement-mapping factor, which captures nonlinear coordinate remapping effects.

We begin by defining the linear dark matter power spectrum as
\begin{equation}
    \langle \widetilde{\delta}^{(1)}(\boldsymbol{k}) \widetilde{\delta}^{(1)}(\boldsymbol{k}') \rangle 
    = (2\pi)^3 \delta_{\rm D}(\boldsymbol{k} + \boldsymbol{k}')\, P_{\rm lin}(\boldsymbol{k}), \label{eq:P_lin_def}
\end{equation}
where the subscript ``lin'' denotes linear theory.

In our formulation, the exponent function $\Sigma_{\rm s,rec}$ and the source correlation function $\xi_{\rm Jgs,rec}$ are expanded perturbatively as
\begin{equation}
\Sigma_{\rm s,rec} = \sum_{n=1}^\infty \Sigma^{(n)}_{\rm s,rec}, \quad
\xi_{\rm Jgs,rec} = \sum_{n=1}^\infty \xi^{(n)}_{\rm Jgs,rec}, \label{eq:sigma_xi_expansions}
\end{equation}
with $\Sigma^{(n)}_{\rm s,rec} = \mathcal{O}(P_{\rm lin}^n)$ and $\xi^{(n)}_{\rm Jgs,rec} = \mathcal{O}(P_{\rm lin}^n)$. The term with $n=1$ corresponds to the tree-level (i.e., linear-order) contribution, while terms with $n \geq 2$ represent the $(n-1)$-loop contributions.

To compute the power spectrum up to one-loop order in our formalism, it suffices to evaluate the following expression:
\begin{align}
    P_{\rm gs,rec}(\boldsymbol{k}) &= e^{-\overline{\Sigma}^{(\rm tree)}_{\rm s,rec}(\boldsymbol{k})}
\int d^3r_q\, e^{-i \boldsymbol{k} \cdot \boldsymbol{r}_q}\,
e^{\Sigma^{(\rm tree)}_{\rm s,rec}(\boldsymbol{k}, \boldsymbol{r}_q)} \nonumber \\
&\quad \times
\left[\xi^{(\rm tree)}_{\rm gs}(\boldsymbol{r}_q) + \xi^{(1\text{-loop})}_{\rm Jgs,rec}(\boldsymbol{r}_q) \right], \label{eq:P_1loop_form}
\end{align}
where the tree-level source correlation function $\xi^{(\rm tree)}_{\rm Jgs}$ does not carry the ``J'' and ``rec'' subscripts, since the displacement field does not enter at linear order.

At first glance, computing $\xi^{(1\text{-loop})}_{\rm Jgs,rec}$ may appear to require a direct evaluation of Eq.~\eqref{eq:xi_J_def}, which could involve cumbersome calculations. However, this is not necessary. Since our formulation is designed to reproduce SPT result order by order, we can deduce $\xi^{(1\text{-loop})}_{\rm Jgs,rec}$ by comparing our expansion to the known one-loop solution in SPT.

In fact, Ref.~\cite{Sugiyama:2024ggt} already provides the one-loop power spectrum incorporating all relevant effects, including redshift-space distortions, galaxy bias, reconstruction, and even discrete sampling corrections arising from the reconstruction process.

Specifically, from Eq.~\eqref{eq:P_expanded}, the following identity holds at one-loop order:
\begin{align}
P^{(1\text{-loop})}_{\rm Jgs,rec}(\boldsymbol{k}) &=
P^{(1\text{-loop})}_{\rm gs,rec}(\boldsymbol{k})
+ \Sigma^{(\rm tree)}_{\rm s,rec}(\boldsymbol{k})\,
P^{(\rm tree)}_{\rm gs}(\boldsymbol{k}) \nonumber \\
&\quad - \int_{\boldsymbol{k}_1 + \boldsymbol{k}_2 = \boldsymbol{k}}
\Sigma^{(\rm tree)}_{\rm s,rec}(\boldsymbol{k}, \boldsymbol{k}_1)\,
P^{(\rm tree)}_{\rm gs}(\boldsymbol{k}_2), \label{eq:P_Jgs_1loop}
\end{align}
where $P^{(\rm tree)}_{\rm gs}$ and $P^{(1\text{-loop})}_{\rm Jgs,rec}$ are the source power spectra corresponding to $\xi^{(\rm tree)}_{\rm gs}$ and $\xi^{(1\text{-loop})}_{\rm Jgs,rec}$, respectively. By computing the right-hand side and applying an inverse Fourier transform, we can thus obtain $\xi^{(1\text{-loop})}_{\rm Jgs,rec}$ without directly calculating Eq.~\eqref{eq:xi_J_def}.

\subsection{IR Cancellation}

Infrared (IR) effects refer to nonlinear contributions arising from large-scale modes with wavenumbers \( p \ll k \), where \( k \) denotes the characteristic scale of interest. In the so-called IR limit, where IR effects dominate and all other nonlinear contributions are neglected, it is well established that the total IR contributions cancel exactly. This phenomenon is known as \textit{IR cancellation}~\cite{Jain:1995kx,Scoccimarro:1995if,Kehagias:2013yd,Peloso:2013zw,Sugiyama:2013pwa,Sugiyama:2013gza,Blas:2013bpa,Blas:2015qsi,Lewandowski:2017kes}.

For example, consider an exact solution of SPT truncated at a given order. If the IR limit is taken for all higher-order terms beyond that truncation, the contributions from these terms cancel out completely, leaving only the truncated SPT solution. In the simplest case, where all nonlinear effects beyond linear order are regarded as IR contributions, IR cancellation implies that only the linear power spectrum remains in the IR limit.

In this subsection, we demonstrate how IR cancellation is realized within the ULPT framework. For simplicity, we focus on the case of dark matter in real space; accordingly, subscript labels such as ``g'' or ``s'' do not appear in this subsection. However, as repeatedly emphasized throughout this paper, our framework treats RSD, galaxy bias, and reconstruction effects in a unified manner. Therefore, the same conclusion holds even when these additional effects are included.

\subsubsection{Overview of Previous Studies}

An intuitive derivation of IR cancellation can be obtained by invoking translational invariance of statistical quantities. In the IR limit, the dominant nonlinear contributions to the dark matter density field arise from long-wavelength displacements, commonly referred to as shift terms. Summing these contributions yields~\cite{Sugiyama:2013gza,Sugiyama:2024eye}
\begin{equation}
    \delta(\boldsymbol{x}) \xrightarrow[]{\text{IR}} 
    \delta^{(1)}(\boldsymbol{x} - \overline{\boldsymbol{\Psi}}^{(1)}),
\label{eq:IR_shift}
\end{equation}
where $\overline{\boldsymbol{\Psi}}^{(1)}$ denotes the linear displacement field evaluated at a fixed position (e.g., the origin), and is effectively constant over the region of interest due to its long-wavelength nature; that is, we define $\overline{\boldsymbol{\Psi}}^{(1)} \equiv \boldsymbol{\Psi}^{(1)}(\boldsymbol{x}=\boldsymbol{0})$. The symbol \( \xrightarrow{\text{IR}} \) indicates the operation of taking the IR limit.

In this limit, IR effects manifest as a uniform coordinate shift of the linear density field. As a result, two-point statistics become
\begin{align}
\langle \delta(\boldsymbol{x}) \delta(\boldsymbol{x}') \rangle
&\xrightarrow[]{\text{IR}} \langle \delta^{(1)}(\boldsymbol{x} - \overline{\boldsymbol{\Psi}}^{(1)}) \delta^{(1)}(\boldsymbol{x}' - \overline{\boldsymbol{\Psi}}^{(1)}) \rangle \nonumber \\
&\xrightarrow[]{\text{IR}} \langle \delta^{(1)}(\boldsymbol{x}) \delta^{(1)}(\boldsymbol{x}') \rangle,
\label{eq:IR_cancel}
\end{align}
where we have assumed that the long-wavelength displacement $\overline{\boldsymbol{\Psi}}^{(1)}$ is statistically uncorrelated with the linear density field $\delta^{(1)}$ in the IR limit. This confirms that, in the IR limit, all IR contributions cancel nonperturbatively, leaving only the linear two-point function.

In Fourier space, Eq.~\eqref{eq:IR_shift} becomes
\begin{equation}
    \widetilde{\delta}(\boldsymbol{k}) \xrightarrow{\rm IR} e^{-i \boldsymbol{k} \cdot \overline{\boldsymbol{\Psi}}^{(1)}} \widetilde{\delta}^{(1)}(\boldsymbol{k}),
\label{eq:IR_fourier_transform}
\tag{164}
\end{equation}
indicating that, in the IR limit, the dominant nonlinear contribution appears as a uniform phase shift of the linear density field.

Using this expression, the power spectrum is computed as
\begin{align}
\langle \widetilde{\delta}(\boldsymbol{k}) \widetilde{\delta}(\boldsymbol{k}') \rangle &\xrightarrow{\rm IR}
\left\langle e^{-i \boldsymbol{k} \cdot \overline{\boldsymbol{\Psi}}^{(1)} - i \boldsymbol{k}' \cdot \overline{\boldsymbol{\Psi}}^{(1)}} \right\rangle
\langle \widetilde{\delta}^{(1)}(\boldsymbol{k}) \widetilde{\delta}^{(1)}(\boldsymbol{k}') \rangle \notag \\
&= (2\pi)^3 \delta_{\rm D}(\boldsymbol{k} + \boldsymbol{k}') \, e^{-\overline{\Sigma}^{\rm (tree)}(\boldsymbol{k})} e^{\overline{\Sigma}^{\rm (tree)}(\boldsymbol{k})} P_{\rm lin}(k) \notag \\
&= (2\pi)^3 \delta_{\rm D}(\boldsymbol{k} + \boldsymbol{k}') P_{\rm lin}(k),
    \label{eq:IR_cancel_power}
\tag{165}
\end{align}
where the variance of the long-wavelength displacement field is given by
\begin{equation}
    \overline{\Sigma}^{\rm (tree)}(\boldsymbol{k}) = \int \frac{d^3 \boldsymbol{k}'}{(2\pi)^3}
\left( \frac{\boldsymbol{k} \cdot \boldsymbol{k}'}{k'^2} \right)^2 P_{\rm lin}(k').
\label{eq:IR_variance}
\tag{166}
\end{equation}
This result confirms that IR cancellation is correctly realized in the power spectrum as well. 

\subsubsection{ULPT-Based Derivation}

To examine the behavior of long-wavelength displacement fields at the field level, we consider a simplifying approximation in the ULPT expression given by Eq.~\eqref{eq:d_dJ_Psi}. Specifically, we evaluate the displacement vector at a fixed spatial position, namely, the origin, and neglect its spatial derivatives. That is, we approximate
\begin{equation}
    \boldsymbol{\Psi}(\boldsymbol{x}) \approx \overline{\boldsymbol{\Psi}}, \quad \text{with} \quad \overline{\boldsymbol{\Psi}} \equiv \boldsymbol{\Psi}(\boldsymbol{x} = 0).
\end{equation}
Under this approximation, Eq.~\eqref{eq:d_dJ_Psi} becomes
\begin{align}
\delta(\boldsymbol{x}) & \approx
\delta_{\rm J}(\boldsymbol{x}) + \sum_{n=1}^{\infty} \frac{(-1)^n}{n!} \overline{\Psi}_{i_1} \cdots \overline{\Psi}_{i_n} \partial_{i_1} \cdots \partial_{i_n} \delta_{\rm J}(\boldsymbol{x}) \nonumber \\
& = \delta_{\rm J}(\boldsymbol{x} - \overline{\boldsymbol{\Psi}}).
\end{align}
This result shows that the contribution from a long-wavelength displacement vector appears as a uniform coordinate shift of the Jacobian deviation field. Substituting the linear-order approximations $\delta_{\rm J} \approx \delta_{\rm J}^{(1)}$ and $\overline{\boldsymbol{\Psi}} \approx \overline{\boldsymbol{\Psi}}^{(1)}$ into this expression reproduces Eq.~\eqref{eq:IR_shift}.

At the power spectrum level, to explicitly observe IR cancellation within the ULPT framework, we consider the simplest setting in which both the displacement-mapping exponent $\Sigma$ and the source correlation function $\xi_{\rm J}$ are evaluated at linear order. That is, we set
\begin{align}
\Sigma(\boldsymbol{k}, \boldsymbol{r}_q) = \Sigma^{(\mathrm{tree})}(\boldsymbol{k}, \boldsymbol{r}_q), \quad
\xi_{\rm J}(\boldsymbol{r}_q) = \xi^{(\mathrm{tree})}(\boldsymbol{r}_q),
\end{align}
and substitute these expressions into the general expression for the power spectrum given in Eq.~\eqref{eq:P_general}.

In this case, the power spectrum is expressed as
\begin{equation}
    P(\boldsymbol{k}) = e^{-\overline{\Sigma}^{(\rm tree)}(\boldsymbol{k})}\int d^3r_q\, e^{-i\boldsymbol{k}\cdot\boldsymbol{r}_q}\, e^{\Sigma^{(\rm tree)}(\boldsymbol{k},\boldsymbol{r}_q)} \xi^{(\rm tree)}(r_q),
    \label{eq:P_IR_tree}
\end{equation}
and its Fourier-space representation reads
\begin{align}
    P(\boldsymbol{k}) & = e^{-\overline{\Sigma}^{(\rm tree)}(\boldsymbol{k})}
     \sum_{n=0}^{\infty}\frac{1}{n!}\int_{\boldsymbol{k}_{1 \cdots (n+1)}=\boldsymbol{k}} \nonumber \\
     & \left[\prod_{i=1}^{n} \left( \frac{\boldsymbol{k}\cdot\boldsymbol{k}_i}{k_i^2} \right)^2 P_{\rm lin}(k_i)  \right] 
     P_{\rm lin}(k_{n+1}),
     \label{eq:IR_cancel_P}
\end{align}
where the scale-dependent part of the displacement-mapping factor is given by
\begin{equation}
    \Sigma^{(\rm tree)}(\boldsymbol{k},\boldsymbol{r}_q) = \int \frac{d^3k'}{(2\pi)^3}
    e^{i\boldsymbol{k}'\cdot\boldsymbol{r}_q} \left( \frac{\boldsymbol{k}\cdot\boldsymbol{k}'}{k'^2} \right)^2 P_{\rm lin}(k').
    \label{eq:sigma_tree}
\end{equation}

In Eq.~\eqref{eq:IR_cancel_P}, the wavevector $\boldsymbol{k}_{n+1}$ corresponds to the contribution from the source power spectrum, while the remaining wavevectors $(\boldsymbol{k}_1, \dots, \boldsymbol{k}_n)$ encode the nonlinear modulation due to long-wavelength displacements.

In the IR limit, where all $\boldsymbol{k}_i$ $(i = 1,\dots,n)$ are much smaller than $\boldsymbol{k}_{n+1}$, i.e., only large-scale (infrared) modes are retained, the Dirac delta function in the mode-coupling integral simplifies as
\begin{equation}
    \delta_{\rm D}(\boldsymbol{k} - \boldsymbol{k}_1 - \cdots - \boldsymbol{k}_{n+1})
    \xrightarrow[]{\text{IR}} \delta_{\rm D}(\boldsymbol{k} - \boldsymbol{k}_{n+1}).
\end{equation}
This approximation implies that the coupling between long- and short-wavelength modes vanishes in the IR limit. Physically, this corresponds to the assumption made in Eq.~\eqref{eq:IR_cancel}, where the long-wavelength displacement vector $\overline{\boldsymbol{\Psi}}^{(1)}$ is uncorrelated with the short-wavelength linear density field $\delta^{(1)}$.

As a result, Eq.~\eqref{eq:IR_cancel_P} reduces to
\begin{equation}
    P(\boldsymbol{k}) \xrightarrow[]{\text{IR}}
    e^{-\overline{\Sigma}^{(\rm tree)}(\boldsymbol{k})}
    e^{\overline{\Sigma}^{(\rm tree)}(\boldsymbol{k})}
    P_{\rm lin}(k) = P_{\rm lin}(k),
\end{equation}
which confirms that IR cancellation is explicitly realized within the ULPT framework.

In the context of these calculations, the IR limit corresponds to taking the small-scale limit $r_q \to 0$ in the configuration-space representation of the displacement-mapping factor, $\Sigma(\boldsymbol{k}, \boldsymbol{r}_q)$. This identification stems from the definition of the IR limit as the large-$k$ regime, $k \to \infty$, where the relevant physical scales are much smaller than the characteristic wavelengths of long-wavelength modes. In configuration space, this corresponds to vanishing separation $r_q \to 0$.

\subsection{IR-Resummed Model}
\label{sec:IR_resummed}

IR cancellation is a phenomenon that holds strictly only in the IR limit. In realistic settings, however, this cancellation is not exact. Assuming that the residual IR contributions primarily affect the baryon acoustic oscillation (BAO) feature, one can construct a model that captures the nonlinear damping of the BAO signal while preserving the smooth broadband shape of the power spectrum predicted by SPT in the absence of BAO. This idea forms the basis of the so-called \textit{IR-resummed model}~\cite{Sugiyama:2013gza,Senatore:2014via,Baldauf:2015xfa,Blas:2016sfa,Senatore:2017pbn,Ivanov:2018gjr,Lewandowski:2018ywf,Sugiyama:2020uil,Sugiyama:2024eye}.

In this section, we present a systematic derivation of the IR-resummed model within the ULPT framework.

To construct the model, we begin by decomposing the linear matter power spectrum into two components: a \textit{wiggle} part, which contains only the BAO feature, and a \textit{no-wiggle} part, which describes the smooth broadband shape without BAO:
\begin{equation}
    P_{\rm lin}(k) = P_{\rm w}(k) + P_{\rm nw}(k),
\end{equation}
where the subscripts ``w'' and ``nw'' denote the wiggle and no-wiggle components, respectively.

The corresponding linear correlation function can likewise be decomposed as
\begin{equation}
    \xi^{(\rm tree)}(r) = \xi^{(\rm tree)}_{\rm w}(r) + \xi^{(\rm tree)}_{\rm nw}(r),
    \label{eq:xi_trew_wnw}
\end{equation}
where \( \xi^{(\rm tree)}_{\rm w}(r) \) and \( \xi^{(\rm tree)}_{\rm nw}(r) \) are the inverse Fourier transforms of \( P_{\rm w}(k) \) and \( P_{\rm nw}(k) \), respectively.

This procedure is commonly referred to as the \textit{wiggle--no-wiggle decomposition}.

\subsubsection{Pre-Reconstruction}
\label{sec:IR_resummed_pre}

To demonstrate that the ULPT framework naturally encompasses the IR-resummed model, we once again consider the simplest setting, in which both the displacement-mapping exponent $\Sigma$ and the source correlation function $\xi_{\rm J}$ are evaluated at linear order. We begin with Eq.~\eqref{eq:P_IR_tree}.

At linear order, the scale dependence of the displacement-mapping factor is given by Eq.~\eqref{eq:sigma_tree}, which evaluates to
\begin{align}
    \Sigma^{(\rm tree)}(\boldsymbol{k}, \boldsymbol{r}_q)
    &= \frac{1}{3} k^2 \int \frac{dk'}{2\pi^2}
    \biggl[ j_0(k' r_q) \nonumber \\
        & \quad - 2 \mathcal{L}_2(\hat{\boldsymbol{k}} \cdot \hat{\boldsymbol{r}}_q)\, j_2(k' r_q)
    \biggr] P_{\rm lin}(k').
\end{align}
where \( j_\ell \) is the spherical Bessel function of order \( \ell \), and \( \mathcal{L}_2 \) is the second Legendre polynomial.

For simplicity, we focus on the isotropic contribution and neglect the second (anisotropic) term on the right-hand side.\footnote{For discussions of the anisotropic contribution, see e.g., Ref.~\cite{Chen:2024pyp}.} Under this approximation, the scale-dependent factor becomes
\begin{equation}
\Sigma^{(\rm tree)}(\boldsymbol{k}, \boldsymbol{r}_q) \approx k^2 \sigma^2(r_q),
\label{eq:sigma_ks}
\end{equation}
where
\begin{equation}
\sigma^2(r_q) = \frac{1}{3} \int \frac{dk}{2\pi^2} j_0(k r_q) P_{\rm lin}(k),
\end{equation}
and the small-separation limit \( r_q \to 0 \) defines the variance
\begin{equation}
\bar{\sigma}^2 = \sigma^2(r_q = 0).
\end{equation}

Substituting Eqs.~(\ref{eq:xi_trew_wnw}) and (\ref{eq:sigma_ks}) into Eq.~\eqref{eq:P_IR_tree}, the power spectrum becomes
\begin{align}
    P(\boldsymbol{k}) & = e^{-k^2 \bar{\sigma}^2} \int d^3 r_q\, e^{-i \boldsymbol{k} \cdot \boldsymbol{r}_q}\, 
    e^{k^2 \sigma^2(r_q)} \nonumber \\
    & \quad \times\left[ \xi_{\rm w}^{(\rm tree)}(r_q) + \xi_{\rm nw}^{(\rm tree)}(r_q)\right].
\end{align}

The first term, \( \xi^{(\rm tree)}_{\rm w}(r_q) \), contains only the BAO feature and exhibits a peak near the BAO scale, \( r_{\rm BAO} \sim 110\, h^{-1} \mathrm{Mpc} \), while being negligible elsewhere. Owing to this localized structure, similar to a Dirac delta function, the scale-dependent damping factor can be approximated as \( \sigma^2(r) \approx \sigma^2(r_{\rm BAO}) \). Under this approximation, the first term becomes
\begin{align}
& e^{-k^2 \bar{\sigma}^2}
\int d^3 r_q\, e^{-i \boldsymbol{k} \cdot \boldsymbol{r}_q}\, e^{k^2 \sigma^2(r_{\rm BAO})} \xi^{(\rm tree)}_{\rm w}(r_q)
\nonumber \\
& = e^{-k^2 \bar{\sigma}^2_{\rm BAO}} P_{\rm w}(k),
\end{align}
where
\begin{align}
\bar{\sigma}^2_{\rm BAO} & = \bar{\sigma}^2 - \sigma^2(r_{\rm BAO}) \nonumber \\
& = \frac{1}{3}\int \frac{dk}{2\pi^2} \left[ 1 - j_0(kr_{\rm BAO})\right]\,P_{\rm lin}(k).
\end{align}

For the second term, \( \xi^{(\rm tree)}_{\rm nw}(r_q) \), we apply the IR limit and approximate \( \sigma^2(r) \approx \bar{\sigma}^2 \), yielding
\begin{align}
    e^{-k^2 \bar{\sigma}^2} \int d^3 r_q\, e^{-i \boldsymbol{k} \cdot \boldsymbol{r}_q}\, e^{k^2 \bar{\sigma}^2} \xi^{(\rm tree)}_{\rm nw}(r_q) 
    = P_{\rm nw}(k).
\end{align}

Combining both contributions, we obtain the well-known tree-level IR-resummed model:
\begin{equation}
    P(k) = e^{-k^2 \bar{\sigma}^2_{\rm BAO}} P_{\rm w}(k) + P_{\rm nw}(k).
\end{equation}

In this model, the broadband shape of the power spectrum is described by the no-wiggle linear component \( P_{\rm nw}(k) \), while the nonlinear damping of the BAO feature is captured by a single Gaussian damping factor applied to the wiggle part \( P_{\rm w}(k) \).

If one wishes to include one-loop corrections, both the wiggle and no-wiggle components receive higher-order modifications. The broadband shape can be modeled using the one-loop SPT no-wiggle spectrum, while the nonlinear evolution of the BAO feature is captured by applying the one-loop correction to the wiggle part. This can be achieved by evaluating the source correlation function \( \xi_{\rm J} \) up to one-loop order and subsequently decomposing it into its wiggle and no-wiggle components.

\subsubsection{Post-Reconstruction}

Here, we derive the tree-level IR-resummed model for the reconstructed power spectrum within the ULPT framework, and explicitly demonstrate that the BAO feature in the post-reconstruction power spectrum is described by a single-Gaussian-type damping factor.

If we restrict the source correlation function to tree level, it remains unaffected by the reconstruction procedure:
\begin{eqnarray}
    \xi^{(\rm tree)}_{\rm J,rec}(r_q) = \xi^{(\rm tree)}(r_q).
\end{eqnarray}
Therefore, we only need to consider the effect of reconstruction on the displacement-mapping factor.

At linear order, the reconstructed displacement field is given by
\begin{equation}
   \boldsymbol{\Psi}_{\rm rec}^{(1)}(\boldsymbol{q})
    = i\,\int \frac{d^3k}{(2\pi)^3} e^{i\boldsymbol{k}\cdot\boldsymbol{q}}
    \frac{\boldsymbol{k}}{k^2}\left[ 1 - W_{\rm G}(kR) \right]\, \widetilde{\delta}^{(1)}(\boldsymbol{k}),
\end{equation}
where \( W_{\rm G}(kR) \) denotes the Gaussian smoothing filter used in reconstruction.

Following the same logic as in Sec.~\ref{sec:IR_resummed_pre}, the scale dependence of the displacement-mapping factor after reconstruction is characterized by
\begin{equation}
    \Sigma_{\rm rec}^{(\rm tree)}(\boldsymbol{k},\boldsymbol{r}_q)
    \approx k^2\sigma_{\rm rec}^2(r_q),
\end{equation}
where
\begin{align}
    \sigma^2_{\rm rec}(r_q) & =
    \frac{1}{3} \int \frac{dk}{2\pi^2} j_0(kr_q)
    \left[ 1 - W_{\rm G}(kR) \right]^2\,P_{\rm lin}(k).
\end{align}
Taking the small-separation limit \( r_q \to 0 \), we define
\begin{eqnarray}
    \bar{\sigma}^2_{\rm rec} = \sigma^2_{\rm rec}(r_q=0).
\end{eqnarray}

By decomposing the tree-level source correlation function \( \xi^{(\rm tree)} \) into wiggle and no-wiggle components, we can derive the IR-resummed model for the reconstructed power spectrum in a form analogous to the pre-reconstruction case~\cite{Sugiyama:2024eye,Chen:2024tfp}:
\begin{equation}
    P(k) = e^{-k^2 \bar{\sigma}^2_{\rm BAO, rec}} P_{\rm w}(k) + P_{\rm nw}(k),
\end{equation}
where
\begin{align}
    \bar{\sigma}^2_{\rm BAO,rec}
    &= \bar{\sigma}^2_{\rm rec} - \sigma^2_{\rm rec}(r_{\rm BAO})  \nonumber \\
    &= \frac{1}{3}\int \frac{dk}{2\pi^2} \left[ 1 - j_0(kr_{\rm BAO})\right] \nonumber \\
    & \quad \times
    \left[ 1 - W_{\rm G}(kR) \right]^2\,P_{\rm lin}(k).
\end{align}

The only difference from the pre-reconstruction case lies in the value of the exponential damping scale \( \bar{\sigma}^2_{\rm BAO,rec} \), which is reduced due to the presence of the multiplicative factor \( \left[ 1 - W_{\rm G}(kR) \right]^2 \) in the integrand. As a result, the nonlinear damping of the BAO feature becomes weaker after reconstruction, and the amplitude of the BAO signal is effectively enhanced compared to the pre-reconstruction case.

\subsection{Cross Spectrum of Pre- and Post-Reconstruction Fields}
\label{sec:cross_P}

It is well established that the cross-power spectrum between pre- and post-reconstruction density fields exhibits an overall exponential damping behavior~\cite{Wang:2022nlx}. This effect originates from the mismatch in IR contributions before and after reconstruction~\cite{Sugiyama:2024eye}. In this subsection, we investigate how this characteristic feature arises naturally within the ULPT framework.

\subsubsection{Overview of Previous Studies}

In the IR limit, the reconstructed dark matter density fluctuation can be approximated as~\cite{Sugiyama:2024eye}
\begin{equation}
    \delta_{\rm rec}(\boldsymbol{x}) \xrightarrow[]{\text{IR}} \delta^{(1)}(\boldsymbol{x} 
    - \overline{\boldsymbol{\Psi}}^{(1)}_{\rm rec}),
\end{equation}
where $\overline{\boldsymbol{\Psi}}^{(1)}_{\rm rec} = \overline{\boldsymbol{\Psi}}^{(1)} + \overline{\boldsymbol{t}}^{(1)}$ denotes the large-scale displacement vector after reconstruction, evaluated at a fixed reference point (typically the origin). In particular, the reconstruction-induced displacement is given at linear order by
\begin{equation}
    \overline{\boldsymbol{t}}^{(1)} = i \int \frac{d^3k}{(2\pi)^3} \frac{\boldsymbol{k}}{k^2} \left[-W_{\rm G}(kR)\right] \tilde{\delta}^{(1)}(\boldsymbol{k}).
\end{equation}

The corresponding cross-correlation function between pre- and post-reconstruction fields is given by
\begin{align}
    \langle \delta_{\rm rec}(\boldsymbol{x})\, \delta(\boldsymbol{x}') \rangle
    &\xrightarrow[]{\text{IR}} \langle \delta^{(1)}(\boldsymbol{x} - \overline{\boldsymbol{\Psi}}^{(1)}_{\rm rec})\, \delta^{(1)}(\boldsymbol{x}' - \overline{\boldsymbol{\Psi}}^{(1)}) \rangle \notag \\
    &\xrightarrow[]{\text{IR}} \langle \delta^{(1)}(\boldsymbol{x} - \overline{\boldsymbol{t}}^{(1)})\, \delta^{(1)}(\boldsymbol{x}') \rangle.
\end{align}
Here, $\delta_{\rm rec}$ and $\delta$ are associated with different large-scale displacement fields. As a result, the translational invariance that normally ensures IR cancellation no longer holds, and a residual contribution from the reconstruction-induced displacement $\overline{\boldsymbol{t}}^{(1)}$ remains.

In Fourier space, the cross-power spectrum receives an exponential contribution originating from this residual displacement:
\begin{align}
    P_{\rm cross}(k) &= \int d^3r\, e^{-i\boldsymbol{k}\cdot\boldsymbol{r}} \langle \delta^{(1)}(\boldsymbol{x} - \overline{\boldsymbol{t}}^{(1)})\, \delta^{(1)}(\boldsymbol{x}') \rangle \notag \\
    &= \langle e^{-i\boldsymbol{k} \cdot \overline{\boldsymbol{t}}^{(1)}} \rangle\, P_{\rm lin}(k),
\end{align}
where $\boldsymbol{r} = \boldsymbol{x} - \boldsymbol{x}'$ is the separation vector between two Eulerian coordinates. This expression assumes that the long-wavelength displacement field is statistically uncorrelated with the short-wavelength linear density field.

The moment-generating function for $(-i\boldsymbol{k} \cdot \overline{\boldsymbol{t}}^{(1)})$ can be evaluated as
\begin{equation}
    \left\langle e^{-i\boldsymbol{k} \cdot \overline{\boldsymbol{t}}^{(1)}} \right\rangle
    = \exp\left[ -\frac{1}{2}k^2\, \bar{\sigma}_{\rm tt}^2 \right],
\end{equation}
where the variance of the reconstruction-induced displacement is given by
\begin{equation}
    \bar{\sigma}_{\rm tt}^2 = \frac{1}{3} \int \frac{dk}{2\pi^2} \left[ W_{\rm G}(kR) \right]^2 P_{\rm lin}(k).
    \label{eq:sigma_tt}
\end{equation}

We thus arrive at the final expression for the cross-power spectrum:
\begin{equation}
    P_{\rm cross}(k) \xrightarrow[]{\text{IR}} e^{-k^2 \bar{\sigma}_{\rm tt}^2 / 2}\, P_{\rm lin}(k),
\end{equation}
in agreement with Eq.~(136) of Ref.~\cite{Sugiyama:2024eye}. This result confirms that the mismatch in IR effects between the pre- and post-reconstruction density fields leads to a Gaussian-type exponential suppression in the cross-power spectrum across all scales.

\subsubsection{ULPT-Based Derivation}

Within our framework, the exponent of the displacement-mapping factor at linear order is expressed as
\begin{align}
    & -\overline{\Sigma}_{\rm cross}^{\rm (tree)}(\boldsymbol{k}) + \Sigma_{\rm cross}^{\rm (tree)}(\boldsymbol{k}, \boldsymbol{r}_q) \notag \\
    &\quad = \frac{1}{2} \left\langle \left\{ -i\boldsymbol{k} \cdot \left[ \boldsymbol{\Psi}_{\rm rec}^{(1)}(\boldsymbol{q}) - \boldsymbol{\Psi}^{(1)}(\boldsymbol{q}') \right] \right\}^2 \right\rangle_{\rm c}.
\end{align}
Here, we emphasize that $\overline{\Sigma}_{\rm cross}^{\rm (tree)}(\boldsymbol{k})$ is not defined as the zero-separation limit of $\Sigma_{\rm cross}^{\rm (tree)}(\boldsymbol{k}, \boldsymbol{r}_q)$.

The explicit form of the exponent function is given by
\begin{align}
    \Sigma_{\rm cross}^{\rm (tree)}(\boldsymbol{k}, \boldsymbol{r}_q)
    &= \int \frac{d^3k'}{(2\pi)^3} \, e^{i\boldsymbol{k}' \cdot \boldsymbol{r}_q} \left( \frac{\boldsymbol{k} \cdot \boldsymbol{k}'}{k'^2} \right)^2 \nonumber \\
    &\quad \times \left[ 1 - W_{\rm G}(k'R) \right] \, P_{\rm lin}(k'),
\end{align}
and
\begin{align}
    \overline{\Sigma}_{\rm cross}^{\rm (tree)}(\boldsymbol{k}) = \Sigma_{\rm cross}^{\rm (tree)}(\boldsymbol{k}, \boldsymbol{r}_q = 0)
    + \frac{1}{2}k^2 \bar{\sigma}^2_{\rm tt},
\end{align}
where $\bar{\sigma}^2_{\rm tt}$ is the variance of the reconstruction-induced displacement field, defined in Eq.~\eqref{eq:sigma_tt}.

Assuming the tree-level source correlation function, the cross-power spectrum takes the form
\begin{equation}
    P_{\rm cross}(k)
    = e^{ -\overline{\Sigma}_{\rm cross}^{\rm (tree)}(\boldsymbol{k}) } \int d^3r_q\, e^{-i\boldsymbol{k} \cdot \boldsymbol{r}_q} \,
    e^{ \Sigma_{\rm cross}^{\rm (tree)}(\boldsymbol{k}, \boldsymbol{r}_q) } \, \xi^{\rm (tree)}(r_q).
    \label{eq:P_cross}
\end{equation}

In our formulation, taking the IR limit corresponds to the operation $r_q \to 0$ in the displacement-mapping factor. Accordingly, the cross-power spectrum in the IR limit becomes
\begin{align}
    P_{\rm cross}(k)
    &\xrightarrow[]{\rm IR}
    e^{ -\overline{\Sigma}_{\rm cross}^{\rm (tree)}(\boldsymbol{k}) } \, e^{ \Sigma_{\rm cross}^{\rm (tree)}(\boldsymbol{k}, \boldsymbol{r}_q = 0) } \, P_{\rm lin}(k) \notag \\
    &= e^{-k^2 \bar{\sigma}^2_{\rm tt} / 2} \, P_{\rm lin}(k),
\end{align}
which reproduces the expected Gaussian-type suppression in the IR limit.

Following the same logic as in Sec.~\ref{sec:IR_resummed}, we can also derive the tree-level IR-resummed model for the cross-power spectrum. By decomposing the linear correlation function $\xi^{\rm (tree)}$ in Eq.~\eqref{eq:P_cross} into wiggle and no-wiggle components, we obtain
\begin{equation}
    P_{\rm cross}(k) = e^{-k^2 \bar{\sigma}_{\rm tt}^2/2}
    \left[ e^{-k^2 \bar{\sigma}_{\rm BAO,cross}^2} \, P_{\rm w}(k) + P_{\rm nw}(k) \right],
\end{equation}
where the effective damping scale is given by
\begin{align}
    \bar{\sigma}^2_{\rm BAO,cross}
    & = \frac{1}{3} \int \frac{dk}{2\pi^2} \left[ 1 - j_0(k r_{\rm BAO}) \right] \nonumber \\
    &\quad \times \left[ 1 - W_{\rm G}(kR) \right] \, P_{\rm lin}(k).
\end{align}
This expression matches Eq.~(138) of Ref.~\cite{Sugiyama:2024eye}.

\section{Future Prospects}
\label{sec:FP}

The ULPT framework provides a unified description of IR effects across a variety of settings, including real-space dark matter, biased tracers such as galaxies, redshift space, and post-reconstruction density fields. This unification is realized through the concept of the \textit{displacement-mapping effect}, which systematically captures nonlinear coordinate remapping. In particular, ULPT naturally incorporates the IR-resummed model for the power spectrum and accurately characterizes the nonlinear damping of the BAO feature in all of the above cases.

While the ULPT formalism provides a robust theoretical prediction for the damping of BAO features, its ability to accurately describe the broadband shape of the power spectrum beyond the BAO scale remains to be tested. A thorough assessment of this capability requires comparisons with $N$-body simulations and observational data, which we leave for future work. Nevertheless, in this section, we explore the potential impact of nonlinear gravitational evolution, galaxy bias, RSD, and reconstruction on the shape of the power spectrum within the ULPT framework.

Finally, we comment on possible extensions of ULPT to higher-order statistics, such as the bispectrum, as well as to other observables including galaxy-galaxy lensing.

\subsection{Gravitational Nonlinearities of Dark Matter}

The ULPT framework is constructed to exactly reproduce the results of standard perturbation theory (SPT) up to a given truncated order, such as one-loop, while treating all higher-order contributions beyond that order (e.g., two-loop and higher corrections) as arising from the displacement-mapping factor. However, it remains an open question to what extent this decomposition faithfully captures the detailed structure of higher-order contributions. 

Importantly, in this work we have analytically demonstrated that the displacement-mapping factor plays an essential role in accurately describing the nonlinear damping of BAO. Going forward, it will be important to investigate whether this factor also governs the overall broadband shape of the power spectrum, thereby exerting a broader impact beyond BAO modeling. The validity of this structural decomposition, particularly in reproducing the broadband shape, should be further assessed through systematic comparisons with numerical simulations and observational data.

\subsection{Galaxy Bias}
\label{sec:galaxy_bias}

As discussed in detail in Sec.~\ref{sec:interp_J}, the ULPT framework explicitly separates the Jacobian deviation from the displacement-mapping contribution. The Jacobian deviation, which represents the intrinsic part of the density fluctuation, is found to involve only Galileon-type nonlinear effects at second and third order in perturbation theory, specifically: $\mathcal{G}^{(2)}_2$, $\mathcal{G}^{(3)}_3$, and $\Gamma^{(3)}_3$. No other operators appear up to third order. This structure arises naturally from the fact that the ensemble average of the Jacobian deviation vanishes at all orders in perturbation theory.

Within the ULPT framework, biased density fluctuations enter as a linear additive contribution to the Jacobian deviation, i.e., in the form $[\delta_{\rm J} + \delta_{\rm b}]$. Assuming that biased tracers, such as galaxies, inherit the nonlinear gravitational evolution of dark matter, the simplest model posits that $\delta_{\rm b}$, like $\delta_{\rm J}$, consists solely of Galileon-type operators. If biased tracers exhibit additional nonlinear contributions not present in dark matter, such effects should be considered only if this minimal model fails to describe the data.

Under the assumption that biased tracers follow the nonlinear structure of dark matter, the second- and third-order bias fluctuations are expected to be proportional to the corresponding Galileon-type operators,
\begin{align}
    \delta_{\rm b}^{(2)} &\propto \mathcal{G}^{(2)}_2, \\
    \delta_{\rm b}^{(3)} &\propto \{\mathcal{G}^{(3)}_3,\, \Gamma^{(3)}_3\}.
\end{align}
Alternatively, within the ULPT framework one may equivalently assume proportionality to the Jacobian deviation,
\begin{align}
    \delta_{\rm b}^{(2)} &\propto \delta_{\rm J}^{(2)}, \\
    \delta_{\rm b}^{(3)} &\propto \{\delta^{(3)}_{{\rm J},U},\, \delta^{(3)}_{{\rm J},V}\},
\end{align}
where the explicit parametrization of such proportionality factors is left for future work. A notable feature of this Galileon-based description is that local operators such as $[\delta^{(1)}]^2$ and $[\delta^{(1)}]^3$ never appear independently; they are always embedded within Galileon combinations. As a result, purely local bias terms do not arise.

In general, the presence of local bias introduces several complications into the modeling of galaxy bias. First, local bias operators typically generate galaxy density fluctuations whose ensemble average is nonzero. This necessitates the artificial subtraction of a constant offset at the field level to enforce $\langle \delta_{\rm g} \rangle = 0$. Even after such a correction, contributions from local bias operators to the power spectrum remain nonvanishing at $k = 0$, requiring either the explicit subtraction of the $k = 0$ mode or the introduction of an additional parameter, often denoted $P_{\epsilon}$, to absorb this constant as a shot-noise term. These procedures fall under the general framework of \textit{bias renormalization} (e.g., see Ref.~\cite{Desjacques:2016bnm}).

By contrast, the Galileon-based bias model proposed within the ULPT framework naturally satisfies $\langle \delta_{\rm g} \rangle = 0$, preserving the same statistical consistency as dark matter density fluctuations. In other words, this bias parametrization allows for fully renormalization-free predictions, in direct analogy with the case of unbiased dark matter. Indeed, it has already been demonstrated in Ref.~\cite{Assassi:2014fva} that terms constructed solely from Galileon-type operators are inherently free of bias, further supporting the validity of this approach.

Whether this model is consistent with results from $N$-body simulations and observational data remains to be tested. Empirical validation of this proposal should therefore be pursued in future work.

\subsection{RSD}

The ULPT framework enables the computation of power spectra by resumming the infinite series of mode-coupling integrals that arise from the displacement-mapping factor. This is achieved through configuration-space convolution integrals, which naturally implement the nonlinear remapping of density fields. Structurally, this approach is analogous to that employed in CLPT~\cite{Carlson:2012bu}, which also operates in configuration space. However, in contrast to CLPT, ULPT retains the Jacobian deviation explicitly, rather than treating only the displacement vector as the fundamental dynamical variable.

By explicitly separating the Jacobian deviation and the displacement-mapping contribution, ULPT also shares conceptual features with the TNS model~\cite{Taruya:2010mx}. In particular, ULPT naturally reproduces the separation between the Kaiser effect and the velocity field contribution in the exponential prefactor. Nevertheless, it is important to note that the TNS model is formulated in the Eulerian framework and adopts a key approximation: the scale dependence of the velocity contribution in the exponential is neglected when evaluating the convolution integrals. This simplification permits the entire computation to be carried out in Fourier space, in contrast to the configuration-space approach of ULPT.

In summary, ULPT inherits essential structural elements from both CLPT and TNS, while extending their capabilities through a unified and physically transparent treatment of nonlinear coordinate transformations. These connections suggest that ULPT provides a promising and extensible framework for modeling RSD effects in a manner that is both accurate and theoretically consistent.

\subsection{Reconstruction}

The post-reconstruction power spectrum is known to exhibit a slight suppression relative to its pre-reconstruction counterpart~\cite{Wang:2022nlx}. This suppression is believed to originate from a mechanism analogous to that of RSD, specifically the addition of reconstruction-induced contributions to the displacement vector.

If the ULPT framework proves capable of accurately modeling the shape of the power spectrum in the presence of RSD, then it is reasonable to expect that ULPT can also provide a reliable description of the power spectrum under reconstruction, which introduces a structurally similar modification to the displacement field.

\subsection{Applications}

\subsubsection{Bispectrum}

Once the practical utility of the ULPT framework for modeling the power spectrum is firmly established, a natural extension is to develop a consistent theoretical model for higher-order statistics, most notably the bispectrum. As the bispectrum contains cosmological information that is complementary to that encoded in the power spectrum, advancing its theoretical modeling is essential for fully exploiting the information content of large-scale structure.

For instance, the treatment of IR contributions involving products of the form $P_{\mathrm{w}}(k_1)P_{\mathrm{w}}(k_2)$ differs between Refs.~\cite{Blas:2016sfa} and \cite{Sugiyama:2020uil,Sugiyama:2024qsw}, resulting in a subtle yet important discrepancy. The ULPT framework, which incorporates IR effects efficiently through an explicit displacement-mapping factor, may offer a consistent resolution to this issue.

Furthermore, it has been shown that the non-Gaussian component of the bispectrum covariance is significantly reduced after reconstruction. Consequently, when constraining primordial non-Gaussianity using the galaxy bispectrum, the achievable constraints can improve by up to a factor of three compared to the pre-reconstruction case~\cite{Shirasaki:2020vkk}.

Given that ULPT provides a unified modeling framework that naturally extends to the post-reconstruction regime, it is expected to play a key role in the development of accurate theoretical models for the reconstructed bispectrum.

\subsubsection{Galaxy-Galaxy-Lensing}

Galaxy--galaxy lensing (GGL), the cross-correlation between spectroscopically observed galaxy clustering and photometrically measured cosmic shear, has recently emerged as a powerful cosmological probe (see, e.g., Refs.~\cite{Miyatake:2023njf,Sugiyama:2023fzm}, and references therein). In the analysis of GGL, the density-field reconstruction technique can be applied to the spectroscopic galaxy sample. As in the case of galaxy auto-correlations, this reconstruction can reduce statistical uncertainties and potentially tighten constraints on cosmological parameters relative to the pre-reconstruction case.

However, since the reconstruction procedure cannot be applied to the cosmic shear field, modeling the cross-power spectrum between pre- and post-reconstruction fields becomes necessary. As discussed in Sec.~\ref{sec:cross_P}, this cross spectrum acquires an overall exponential suppression arising from residual IR displacements. The ULPT framework developed in this work provides a natural and explicit prediction for this damping factor, and therefore offers a promising theoretical framework for analyzing GGL observables that incorporate reconstructed galaxy samples.

\section{Conclusion}
\label{sec:Conclusion}

We have developed Unified Lagrangian Perturbation Theory (ULPT), a perturbative framework that provides a consistent treatment of galaxy density fluctuations in real space, redshift space, and after reconstruction. By explicitly separating the observed density field into two components—the Jacobian deviation and the displacement-mapping effect—ULPT captures both intrinsic nonlinear evolution and large-scale convective distortions in a structurally transparent way.

This decomposition satisfies the theoretical requirements of infrared (IR) safety. It enables a fully analytic treatment of IR effects, including exact IR cancellation, single-Gaussian damping of the baryon acoustic oscillations (BAO), and the correct residual structure in cross-power spectra. Moreover, ULPT recovers standard perturbation theory at each order while organizing higher-order corrections through the displacement-mapping exponent.

The structure of the Jacobian deviation naturally generates Galileon-type scalar operators at second and third order, forming a compact basis for nonlinear gravitational effects. This, in turn, facilitates a systematic and physically motivated modeling of Lagrangian galaxy bias, which enters additively into the ULPT formulation.

A unified expression for the power spectrum, derived within the ULPT framework, applies equally to dark matter, biased tracers, redshift-space distortions, and reconstructed fields. Future extensions to higher-order statistics, such as the bispectrum, may benefit from the same decomposition strategy. In addition, the framework is expected to have direct applications to other two-point observables, including galaxy--galaxy lensing. Systematic comparisons with simulations and data will be essential to evaluate the full predictive power of ULPT and to explore possible phenomenological refinements.

\begin{acknowledgments}
N.S. acknowledges financial support from JSPS KAKENHI Grant No. 25K07343, administratively hosted by the National Astronomical Observatory of Japan. N.S. also acknowledges the use of \textit{ChatGPT} (OpenAI) for assistance in language refinement and literature exploration during the preparation of this manuscript.
\end{acknowledgments}

\bibliography{ms}

\end{document}